\documentclass[superscriptaddress,aps,pra,10pt,twocolumn,showpacs,notitlepage,nofootinbib,longbibliography]{revtex4-1}
\usepackage{etex}
\usepackage{amsmath,amssymb,amsthm}
\usepackage[colorlinks=true,citecolor=blue,urlcolor=blue]{hyperref}
\usepackage[pdftex]{graphicx}
\usepackage{times,txfonts}
\usepackage{braket}
\usepackage{color}
\usepackage{natbib}
\usepackage{amsmath,blkarray}
\usepackage{mathtools}
\usepackage{latexsym}
\usepackage{tabularx, booktabs}
\usepackage{graphics,epstopdf}
\usepackage{graphicx}
\usepackage{float}
\usepackage{graphicx}
\usepackage{amsfonts}
\usepackage{subcaption}
\usepackage{multirow}

\newcommand{\be}{\begin{equation}}
\newcommand{\ee}{\end{equation}}
\newcommand{\ba}{\begin{eqnarray}}
\newcommand{\ea}{\end{eqnarray}}

\allowdisplaybreaks
\usepackage[dvipsnames]{xcolor}

\begin{document}

\title{Creating quantum correlations in generalized entanglement swapping}

\author{Pratapaditya Bej}
\email{pratap6906@gmail.com }
\affiliation{Department of Physics and Center for Astroparticle Physics and Space Science, Bose Institute, EN Block, Sector V, Saltlake, Kolkata - 700091, India}

\author{Arkaprabha Ghosal}
\email{a.ghosal1993@gmail.com}
\affiliation{Department of Physics and Center for Astroparticle Physics and Space Science, Bose Institute, EN Block, Sector V, Saltlake, Kolkata - 700091, India}

\author{Arup Roy}
\email{arup145.roy@gmail.com}
\affiliation{Department of Physics, A B N Seal College, Cooch Behar, West Bengal 736 101, India}

\author{Shiladitya Mal}
\email{shiladitya.27@gmail.com}
\affiliation{Harish-Chandra Research Institute and HBNI, Chhatnag Road, Jhunsi, Allahabad - 211019, India}
\affiliation{Department of Physics and Center for Quantum Frontiers of Research and Technology (QFort),
National Cheng Kung University, Tainan 701, Taiwan}
\affiliation{Physics Division, National Center for Theoretical Sciences, Taipei 10617, Taiwan}

\author{Debarshi Das}
\email{dasdebarshi90@gmail.com}
\affiliation{S. N. Bose National Centre for Basic Sciences, Block JD, Sector III, Salt Lake, Kolkata 700 106, India}
\affiliation{Department of Physics and Astronomy, University College London, Gower Street, WC1E 6BT London, United Kingdom}

\begin{abstract}

We study how different types of quantum correlations can be established as the consequence of a generalized entanglement swapping protocol where starting from two Bell pairs ($1,2$) and ($3,4$), a general quantum measurement (denoted by a positive operator-valued measure or POVM) is performed on the pair ($2,3$), which results in creating quantum correlation in ($1,4$) shared between two spatially separated observers.  Contingent upon using different kinds of POVMs, we show generation or destruction of different quantum correlations in the pairs ($1,4$), ($1,2$) and ($3,4$). This thus reflects non-trivial transfer of quantum correlations from the pairs ($1,2$) and ($3,4$) to the pair ($1,4$).  
As an offshoot, this study provides an operational tool to generate different types of single parameter families of quantum correlated states (for example, entangled but not EPR steerable, or EPR steerable but not Bell nonlocal, or Bell nonlocal) by choosing different quantum measurements in the basic entanglement swapping setup. We further extend our study by taking mixed initial states shared by the pairs (1,2) and (3,4). Finally, we study network nonlocality in our scenario. Here, we find out appropriate POVM measurement for which the generated correlation demonstrates/does not demonstrate network nonlocality for the whole range of the measurement parameter.
        
\end{abstract}

\maketitle

\section{Introduction}
Entanglement swapping  is a protocol by which quantum systems that have never interacted in the past can become entangled \cite{es1}. This protocol can be described as follows: Alice and Bob share the  pair ($1,2$) of two qubits in the Bell state. Bob and Charlie share another pair ($3,4$) of two qubits in the same Bell state. Bob performs projective measurement in the Bell basis on the pair ($2,3$) and communicates the outcome to Alice and Charlie. Alice and Charlie thus end up with a Bell pair ($1,4$).  This protocol can be considered as the quantum teleportation \cite{tele} of a qubit that is maximally entangled with another qubit. 

{\c In the above standard entanglement swapping protocol (using two initial Bell states, measurements in the Bell basis and classical communication), entanglement is completely destroyed in the pairs (1,2) and (3,4) and maximally entangled state is created in the pair (1,4). In other words, quantum correlation (in the form of entanglement) is transferred  completely from the pairs (1,2) and (3,4) to the pair (1,4).}  The  entanglement swapping protocol can be generalized in various ways- by modifying the initial states, or by modifying the measurement performed by Bob, or by extending the number of parties \cite{sbose,gour}. These types of generalized entanglement swapping protocol have been studied in different contexts ranging from quantum networks \cite{perseguers,perseguers2}, quantum nonlocality  \cite{nonlsw1,nonlsw2} to information loss or gain \cite{lg1,pbej}. Hence, the natural question that arises in this context is
that how to design different types of quantum correlation transfer from (1,2) and (3,4) to (1,4) by generalizing the standard entanglement swapping protocol, i.e.,  either by modifying the initial states, or by modifying the
measurements, or both.

In the present study, we consider a generalized entanglement swapping protocol, where each of the two pairs- Alice-Bob, Bob-Charlie initially shares a two-qubit Bell state and  Bob performs a general quantum measurement. A general quantum measurement is represented by a positive operator-valued measure (POVM) \cite{um1,unsharp_measurement}. POVM has been shown to possess operational advantages in many tasks over projective measurements, for example, in the context of distinguishing non-orthogonal quantum states \cite{povmapp1}, demonstrating hidden nonlocality \cite{povmapp2,povmapp3}, probing temporal correlations \cite{tempdas}, sequential detection of quantum correlations by multiple observers \cite{smal,ddas,prl_brown,ssasmal,abera}, recycling resources in information theoretic tasks \cite{ractt,sroytt,rac_debarshi,mal} and so on. Here, our aim is to investigate how different types of quantum correlations are generated in the pair ($1,4$) and are deteriorated in the pairs ($1,2$) and ($3,4$) depending on the choice of the POVM by Bob.

In particular, we focus on the creation or destruction of the following types of quantum correlations- entanglement \cite{entre}, Einstein-Podolsky-Rosen (EPR) steering \cite{stere} and Bell nonlocality \cite{belre}.  In general, these three types of quantum correlations are inequivalent and there exists a hierarchy between these- Bell-nonlocal states form a strict subset of steerable states which again form a strict subset of entangled states \cite{hie1,hie2}. All these three types of quantum correlations are the building blocks for different types of information theoretic and communication tasks. For example, entanglement is useful as resource in quantum teleportation \cite{tele}, dense coding \cite{dc} and so on. Bell nonlocality is the resource for a number of device-independent tasks, e.g, quantum key distribution \cite{bnap1}, randomness certification \cite{bnap2}, randomness amplification \cite{bnap3}, Bayesian game theory \cite{bnap4}. EPR steering is the resource in one-sided device-independent quantum key distribution \cite{steap1}, secure quantum teleportation \cite{steap2}, quantum communication \cite{steap2}, and one-sided device-independent randomness generation \cite{steap3}. Against the above backdrop, constructions of these quantum correlations are one of the basic requirements of modern quantum technology.  The present article provides with some basic operational tools in generalized entanglement swapping setup for generating these correlations in the pair ($1,4$) shared between two spatially separated observers that have never interacted before. 
Specifically, we report the generation of the following types of two-qubit states in ($1,4$) by choosing appropriate POVMs- (i) single parameter class of mixed states that is Bell nonlocal for the whole range of the state parameter, (ii) single parameter class of  mixed states that is steerable, but not Bell nonlocal for the whole range of the state parameter, (iii) single parameter class of  mixed states that is entangled, but not EPR steerable for the whole range of the state parameter, and finally (iv)  single parameter class of  mixed states that gradually shows all the above-mentioned quantum correlations as the state parameter is varied. 
 We further extend our study by probing the above aspects of creating/destroying quantum correlations in the pair (1,4)  when mixed states are initially shared by the pairs (1,2) and (3,4) and appropriate POVMs are performed on (2,3). One important point to  stress  here is that while probing standard Bell nonlocality or EPR steering of the state shared by (1,4), we consider that the sources producing the two initial entangled states are not necessarily independent.

It is to be noted that Bell-nonlocality has been studied in the context of entanglement swapping scenario earlier \cite{nonlsw2}. However, contrary to the usual Bell scenario, entanglement swapping scenario involves uncorrelated entangled states produced from several independent sources \cite{brain2010,brain2012}.  To characterize nonlocality in this context, local models with  uncorrelated local variables are desired. This is the motivation behind introducing the concept of network nonlocality \cite{brain2010}. Network nonlocality in entanglement swapping network scenario can be checked using what is known as  ``bilocality'' inequality \cite{brain2010,gisin2017,brain2012}. The concept of network nonlocality and bilocality inequality have attracted several attentions in recent years as witnessed from a series of works \cite{brain2010, gisin2017,brain2012,tavakoli2022,andreoli2017,gupta2018,kundu2020}. Recently, Gisin \textit{et al.} \cite{gisin2017} showed that the bilocality inequality can be violated when arbitrary pure entangled states are initially shared by the pairs (1,2) and (3,4) and when projective measurement in the Bell basis is performed on (2,3). Further, when two mixed states $\rho_{12}$ and $\rho_{34}$ is initially shared between the pairs (1,2) and (3,4) and  projective measurement in the Bell basis is performed on (2,3), then violation of the bilocality inequality implies that either
$\rho_{12}$ or $\rho_{34}$ or both are Bell-nonlocal \cite{gisin2017}. Motivated from these studies, we extend our study for probing network nonlocality in the generalized entanglement swapping scenario where two two-qubit Bell states are initially shared by the pairs (1,2) and (3,4) and a POVM is performed on (2,3). Here we find out appropriate POVM for which the resulting measurement statistics violates or does not violate the bilocality inequality for the whole range of the measurement parameter. To the best of our knowledge, the present study for the first time in the literature probes network nonlocality under POVM. 

The rest of the paper is organized as follows. In Section \ref{sec2}, we present the mathematical tools that are useful for the present study. In Section \ref{sec3}, we describe in details the generalized entanglement swapping scenario considered by us. Next, in Section \ref{sec4}, we demonstrate our main findings with different classes of POVMs. Finally, we conclude with a short discussion in  Section \ref{sec5}.

\section{Preliminaries } \label{sec2}
In this section, we discuss in brief the quantifications of Bell-nonlocality, EPR steering, entanglement pertaining to two-qubit states that will be used as tools in the present study. 

\subsection{Quantification of Bell-nonlocality}

Whether an arbitrary two-qubit state $\rho$ violates the Bell-CHSH inequality \cite{bell,CHSH} or not is determined by the function  $M_{\rho}$ defined as \cite{horo,horo1}
\begin{equation}
 \label{18}
 M_{\rho}=\max\limits_{i>j}\{t_i+t_j\}.
\end{equation}
Here $t_i$ ($i \in \{1,2,3\}$) are the eigenvalues of the real symmetric matrix $T_{\rho}^TT_{\rho}$, where $T_{\rho}$ is a real $3 \times 3$ matrix with elements $T_{ij} = \text{Tr}[\rho \, (\sigma_i \otimes \sigma_j)]$ ($i, j \in \{1,2,3\}$); $T_{\rho}^T$ is the  transpose of $T_{\rho}$; $\sigma_i$ with $i \in \{1,2,3\}$ are the Pauli matrices. The function $M_{\rho}$ is related to the maximal mean value $B_{\rho}$ of the Bell-CHSH operator through the relation \cite{horo,horo1}, $B_{\rho} = 2 \sqrt{M_{\rho}}$. Hence, $M_{\rho} > 1$ implies   $B_{\rho} > 2$- the condition for quantum violation of the Bell-CHSH inequality by the state $\rho$.   One can, therefore, quantify the degree of Bell-nonlocality pertaining to the two-qubit state $\rho$ using  $\mathcal{N}_{\rho}\propto\max{\{0,B_{\rho}-2\}}$ \cite{costa}. After simplifying and taking a appropriate normalization, we have 
\begin{equation}
 \mathcal{N}_{\rho}=\max{\left\{0,\frac{\sqrt{M_{\rho}}-1}{\sqrt{2}-1}\right\}}.
 \label{bbq}
 \end{equation}
 Consequently, $\mathcal{N}_{\rho} > 0$ implies that the Bell-CHSH inequality is violated by the two-qubit state $\rho$. In the present study, Bell nonlocality of an two-qubit state is probed using the above measure.

\subsection{Quantification of EPR steering}

Similar to the case of Bell-nonlocality, EPR steering of a two-qubit state $\rho$ can be quantified \cite{costa} in terms of the maximum quantum violation of the $n$-setting linear steering inequality proposed in \cite{CJWR}. For $n=2$, the analytical form of this quantifier is given by \cite{costa}, 
\begin{align}
  &\mathcal{S}^{(2)}_{\rho}=\max{\left\{0,\frac{\sqrt{\varLambda^{(2)}_{\rho}}-1}{\sqrt{2}-1}\right\}} \nonumber \\
  & \text{with} \, \,  \, \, \varLambda^{(2)}_{\rho} = \max\limits_{i>j}\{t_i+t_j\},
\end{align}
with $t_i$ ($i \in \{1,2,3\}$) being defined earlier. Further, $\mathcal{S}^{(2)}_{\rho} > 0$ implies that the $2$-setting linear steering inequality \cite{CJWR} is violated by the two-qubit state $\rho$ \cite{costa}.

Next, $n=3$, the analytical form of this quantifier is given by \cite{costa}, 
\begin{align}
  &\mathcal{S}^{(3)}_{\rho}=\max{\left\{0,\frac{\sqrt{\varLambda^{(3)}_{\rho}}-1}{\sqrt{3}-1}\right\}} \nonumber \\
  & \text{with} \, \,  \, \, \varLambda^{(3)}_{\rho} = t_1+t_2+t_3.
  \label{ssq}
\end{align}
 With this, $\mathcal{S}^{(3)}_{\rho} > 0$ implies that the $3$-setting linear steering inequality \cite{CJWR} is violated by the two-qubit state $\rho$ \cite{costa}.
 
 One can easily check that for any two-qubit state $\rho$, $\mathcal{S}^{(2)}_{\rho}=\mathcal{N}_{\rho}$ \cite{girdhar}. Also, a two-qubit state $\rho$ violates the $2$-setting linear steering inequality if and only if the state violates the Bell-CHSH inequality \cite{girdhar}. Hence, non-equivalance of EPR steering and Bell-nonlocality can be probed if we quantify Bell-nonlocality by $\mathcal{N}_{\rho}$ and EPR steering by $\mathcal{S}^{(3)}_{\rho}$. This motivates us to use $\mathcal{S}^{(3)}_{\rho}$ as the quantifier of EPR steering pertaining to any two-qubit state $\rho$. In what follows, for simplicity, we will denote  $\mathcal{S}^{(3)}_{\rho}$ by $\mathcal{S}_{\rho}$. We will determine whether a two-qubit state $\rho$ is steerable or not depending on whether $\mathcal{S}_{\rho} > 0$ or $\mathcal{S}_{\rho} \leq 0$ respectively.

\subsection{Quantification of entanglement}
Entanglement of a two-qubit state $\rho$ can be quantified using the concept of negativity defined as \cite{ppt,vidal}
\begin{equation}
\label{4}
 \mathcal{E}_{\rho}=\|\rho^{T_B}\|_1-1 \nonumber 
\end{equation}
where $\|\rho^{T_B}\|_1$ denotes the trace norm of the partial transpose of $\rho$. From the above definition, we have 
\begin{equation}
 \mathcal{E}_{\rho}=2\max{(0,-\mu_4)}   
 \label{eeq}
\end{equation}
where $\mu_4$ is the minimum eigenvalue of $\rho^{T_{B}}$. A two-qubit state $\rho$ is entangled if and only if $\mathcal{E}_{\rho} > 0$.

\subsection{Positive operator-valued measure (POVM)}

In general, any quantum measurement is represented by POVM \cite{um1,unsharp_measurement}, which is a set of positive operators that add to identity, i.e., $E\equiv \lbrace E_{i}\vert\sum E_{i}=\mathbb{I},0< E_i\leq \mathbb{I}\rbrace$. Every $E_i$ can be  decomposed as $E_i=M_i^{\dagger}M_i$. The effects ($E_{i}s$) represent quantum events that may occur as outcomes of a measurement. For an arbitrary state $\rho$, the post-measurement state after obtaining the outcome $i$ is given by, $\rho_{\text{PM}|i} = \frac{M_i\rho M_i^{\dagger}}{Tr(\rho E_i)} = \frac{U_i \sqrt{E_i}\rho \sqrt{E_i}^{\dagger} U_i^{\dagger}}{Tr(\rho E_i)}$ where $U_i$ is an arbitrary unitary operator (specified by the actual physical realization of the POVM) and the probability of getting the $i$-th outcome is given by $\mbox{Tr}(\rho E_{i})$. 	
As a special case, the post-measurement state, when the $i$-th outcome is obtained, can be deterimined using the generalized von Neumann-L\"{u}ders transformation rule \cite{um1,unsharp_measurement} as follows,
$\rho_{\text{PM}|i} = \frac{\sqrt{E_{i}} \, \rho \, \sqrt{E_{i}}^{\dagger}}{\text{Tr}(\rho E_{i})}$. In the present study, the post-measurement states will be evaluated using the above-mentioned generalized von Neumann-L\"{u}ders transformation rule.

\subsection{Network Nonlocality}

Next, we are going to  describe in brief the basic concept and definition of bilocality and network nonlocality \cite{brain2010}. In a typical entanglement swapping scenario involving three observers, say, Alice, Bob and Charlie, there are two independent sources $S_1$ and $S_2$. Bob shares two pairs of particles, one with Alice produced from  $S_1$ and  another with Charlie produced from  $S_2$. 

In order to capture nonlocality in the above network, let Alice, Bob and Charlie receive inputs denoted by $\Tilde{x}$, $\Tilde{y}$, $\Tilde{z}$ respectively  and return outputs denoted by $\Tilde{a}$, $\Tilde{b}$ and $\Tilde{c}$ respectively. The produced correlation $P(\Tilde{a},\Tilde{b},\Tilde{c}|\Tilde{x},\Tilde{y},\Tilde{z})$ will exhibit network nonlocality if it cannot be described by the following bilocal model \cite{brain2010},
\begin{eqnarray}
    \label{62}
    &&P(\Tilde{a},\Tilde{b},\Tilde{c}|\Tilde{x},\Tilde{y},\Tilde{z})=\nonumber\\
    &&\int d\Lambda_1 d\Lambda_2 \rho(\Lambda_1,\Lambda_2)P(\Tilde{a}|\Tilde{x},\Lambda_1) P(\Tilde{b}|\Tilde{y},\Lambda_1,\Lambda_2)P(\Tilde{c}|\Tilde{z},\Lambda_2)\nonumber\\
    &&\text{with} \hspace{0.2cm} \rho(\Lambda_1,\Lambda_2) = \rho_1(\Lambda_1) \rho_2(\Lambda_2),
\end{eqnarray}
where $\Lambda_1$ and $\Lambda_2$ are two hidden variables produced by two sources $S_1$ and $S_2$ respectively. Here the condition $\rho(\Lambda_1,\Lambda_2) =\rho_1(\Lambda_1) \rho_2(\Lambda_2)$ implies that the two sources $S_1$ and $S_2$ are independent.

The first approach towards deriving bilocal inequality was reported in \cite{brain2010} and later Branciard \textit{et al.} derived more general bilocal  inequalities \cite{brain2012}. Here, we will consider the latter bilocal inequalities. Consider the entanglement swapping scenario with  each of Alice and Charlie performing a single particle measurement with binary inputs $\Tilde{x}\in\{0,1\}$, $\Tilde{z}\in\{0,1\}$ respectively and obtaining corresponding binary outputs $\Tilde{a} \in \{0,1\}$, $\Tilde{c}\in\{0,1\}$ respectively. The middle party, Bob,  performs same measurement (hence receives no input $\Tilde{y}$) with four possible outcomes. Let us denote Bob's outcome by two bits- $\Tilde{b}_0 \Tilde{b}_1 =$ $00$, $01$, $10$, or $11$ (for details see \cite{brain2012,tavakoli2022}). Therefore, the tripartite correlations can be written as $P(\Tilde{a},\Tilde{b}^0 \Tilde{b}^1,\Tilde{c}|\Tilde{x},\Tilde{z})$.  With these, let us define the following \cite{brain2012},
\begin{eqnarray}
\label{64}
 I&=&\langle(A_0+A_1)B^0(C_0+C_1)\rangle, 
\end{eqnarray}
and
\begin{eqnarray}
\label{65}
 J&=&\langle(A_0-A_1)B^1(C_0-C_1)\rangle, 
\end{eqnarray}
where 
\begin{align}
    \langle A_x B^y C_z \rangle = \sum_{\Tilde{a}, \Tilde{b}^0 \Tilde{b}^1, c} (-1)^{\Tilde{a} + \Tilde{b}^y + \Tilde{c}} P(\Tilde{a},\Tilde{b}^0 \Tilde{b}^1,\Tilde{c}|\Tilde{x},\Tilde{z})
    \label{etop}
\end{align}
Using these notations, the bilocality inequality \cite{brain2012} can be expressed as 
\begin{equation}
\label{63}
\mathcal{B}_{bilocal}=\sqrt{|I|}+\sqrt{|J|}\leq2
\end{equation} 


When the above inequality is violated, then the tripartite measurement correlation is non-bilocal, i.e.,  network nonlocality is demonstrated \cite{brain2010,gisin2017,brain2012}.


\section{Setting up the Scenario} \label{sec3}
Specifically, we consider a general entanglement swapping protocol, where Alice shares a Bell pair $|\psi_1\rangle_{12}=(1/\sqrt{2})(|00\rangle+|11\rangle)$ denoted by ($1,2$) with Bob and Bob shares another Bell pair $|\psi_1\rangle_{34}=(1/\sqrt{2})(|00\rangle+|11\rangle)$ denoted by ($3,4$) with Charlie. Here, Alice, Bob and Charlie are physically separated from each other. Bob performs a joint quantum measurement with four possible outcomes (which is a POVM $E\equiv \lbrace E_{1}, E_2, E_3, E_4 \rbrace$) on ($2,3$) and discloses the outcome to both Alice and Charlie. Consequently, the pair ($1,4$) shared between Alice and Charlie may become correlated depending on the choice of the POVM by Bob. Similarly, the correlations present in the pairs ($1,2$) and ($3,4$) may get reduced. Within this framework, different types of POVM leads to different types of correlation transfer from the pairs ($1,2$) and ($3,4$) to the pair ($1,4$). 

In the above scenario, the post-measurement state (after Bob performs the POVM on ($2,3$) and discloses the outcome to Alice and Charlie) shared between Alice-Charlie, when the $i$-th outcome ($i \in \{1,2,3,4\}$) is obtained, is given by,  
\begin{widetext}
\begin{equation}
\label{eq6}
 \rho_{14|i}=\frac{\mbox{Tr}_{23} \left[\left(\mathbb{I}_2\otimes\sqrt{E_{i}}\otimes\mathbb{I}_2 \right) \left(|\psi_1\rangle\langle \psi_1|_{12} \otimes |\psi_1\rangle\langle \psi_1|_{34} \right) \left(\mathbb{I}_2\otimes \sqrt{E_{i}}^{\dagger} \otimes\mathbb{I}_2\right) \right]}{\mbox{Tr} \left[\left(\mathbb{I}_2\otimes\sqrt{E_{i}}\otimes\mathbb{I}_2 \right) \left(|\psi_1\rangle\langle \psi_1|_{12} \otimes |\psi_1\rangle\langle \psi_1|_{34} \right) \left(\mathbb{I}_2\otimes \sqrt{E_{i}}^{\dagger} \otimes\mathbb{I}_2\right) \right]}.
\end{equation}

Similarly, the post-measurement state shared between Alice-Bob and Bob-Charlie, when the $i$-th outcome is obtained, are given by,  
\begin{equation}
 \rho_{12|i}=\frac{\mbox{Tr}_{34} \left[\left(\mathbb{I}_2\otimes\sqrt{E_{i}}\otimes\mathbb{I}_2 \right) \left(|\psi_1\rangle\langle \psi_1|_{12} \otimes |\psi_1\rangle\langle \psi_1|_{34} \right) \left(\mathbb{I}_2\otimes \sqrt{E_{i}}^{\dagger} \otimes\mathbb{I}_2\right) \right]}{\mbox{Tr} \left[\left(\mathbb{I}_2\otimes\sqrt{E_{i}}\otimes\mathbb{I}_2 \right) \left(|\psi_1\rangle\langle \psi_1|_{12} \otimes |\psi_1\rangle\langle \psi_1|_{34} \right) \left(\mathbb{I}_2\otimes \sqrt{E_{i}}^{\dagger} \otimes\mathbb{I}_2\right) \right]}
 \label{post12}
\end{equation}
and 
\begin{equation}
 \rho_{34|i}=\frac{\mbox{Tr}_{12} \left[\left(\mathbb{I}_2\otimes\sqrt{E_{i}}\otimes\mathbb{I}_2 \right) \left(|\psi_1\rangle\langle \psi_1|_{12} \otimes |\psi_1\rangle\langle \psi_1|_{34} \right) \left(\mathbb{I}_2\otimes \sqrt{E_{i}}^{\dagger} \otimes\mathbb{I}_2\right) \right]}{\mbox{Tr} \left[\left(\mathbb{I}_2\otimes\sqrt{E_{i}}\otimes\mathbb{I}_2 \right) \left(|\psi_1\rangle\langle \psi_1|_{12} \otimes |\psi_1\rangle\langle \psi_1|_{34} \right) \left(\mathbb{I}_2\otimes \sqrt{E_{i}}^{\dagger} \otimes\mathbb{I}_2\right) \right]}
 \label{post34}
\end{equation}
respectively. Here, the post-measurement states are determined using the Luders state-update rule.
\end{widetext}

Let each $E_i$ admits the following decomposition,
\begin{equation}
    E_i = \sum_{k=0}^{3} q_{ik} |\Psi_{ik}\rangle \langle \Psi_{ik}|, \nonumber 
\end{equation}
where $|\Psi_{ik}\rangle$ for $k=0,1,2,3$ form a complete orthonormal basis on $\mathbb{C}^2 \otimes \mathbb{C}^2$ and $q_{ik} \geq 0$ for all $k=0,1,2,3$. With this, we have \cite{pbej}
\begin{align}
    &\rho_{14|i} = \frac{E_i^{*}}{\mbox{Tr}\left( E_i \right)} \nonumber \\
    & \text{with} \, \, \, E_i^{*}=\sum_{k=0}^{3} q_{ik} |\Psi_{ik}^{*}\rangle \langle \Psi_{ik}^{*}|,
    \label{post14}
\end{align}
with $|\Psi_{ik}^{*}\rangle$ being the complex conjugation of $|\Psi_{ik}\rangle$ in the computational basis.

However, obtaining similar compact expressions for ($1,2$) and ($3, 4$) seems difficult for an arbitrary POVM.

\section{Results} \label{sec4}
We have analyzed with different classes of POVMs. Below, we present some of these results which seem interesting in the context of quantum correlation (e.g., Bell-nonlocality , EPR steering, entanglement) transfer from the pairs ($1,2$) and ($3,4$) to the pair ($1,4$).

\subsection*{Case I}

At first, we consider a POVM $E\equiv \lbrace E_{1}, E_2, E_3, E_4 \rbrace$ with the following effect operators,
\begin{equation}
  E_i=\lambda|\psi_i\rangle\langle \psi_i| + \frac{1-\lambda}{4} \mathbb{I}_4, 
  \label{wernerbasis}
\end{equation}
where $\lambda \in [0,1]$ is the sharpness parameter denoting the measurement strength; $(1-\lambda)$ denotes the amount of white noise incorporated in the measurement; $|\psi_i\rangle$ for $i = 1,2,3,4$ forms the Bell basis defined as
\begin{align}
 &|\psi_{1}\rangle=\frac{|00\rangle + |11\rangle}{\sqrt{2}}, \hspace{0.5cm} |\psi_{2}\rangle=\frac{|00\rangle - |11\rangle}{\sqrt{2}}, \nonumber \\
 &|\psi_{3}\rangle=\frac{|01\rangle + |10\rangle}{\sqrt{2}}, \hspace{0.5cm} |\psi_{4}\rangle=\frac{|01\rangle - |10\rangle}{\sqrt{2}}. \nonumber 
\end{align}
The above measurement becomes projective when $\lambda=1$ and becomes trivial when $\lambda=0$. This measurement is unentangled for $0 \leq \lambda \leq \frac{1}{3}$ and entangled for $\frac{1}{3} < \lambda \leq 1$.

For this POVM, each outcome $i$ occurs with probability $p_i=\frac{1}{4}$ for all $i \in \{1,2,3,4\}$.

After Bob gets the outcome $i$ ($i \in \{1,2,3,4\}$) contingent upon performing the above POVM and communicates the outcome to Alice and Charlie, the post-measurement state shared between Alice and Charlie becomes 
\begin{equation}
 \label{werner14}
 \rho_{14|i}=\lambda|\psi_i\rangle\langle \psi_i| + \frac{1-\lambda}{4} \mathbb{I}_4  \, \, \, \forall i \in \{1,2,3,4\}.
\end{equation}
This is obtained from Eq.(\ref{post14}).

Similarly using the Eqs.(\ref{post12}) and (\ref{post34}), we get the post-measurement state shared between Alice-Bob and that between Bob-Charlie as written below, 
\begin{align}
 \label{werner1234}
 &\rho_{12|i}= \rho_{34|i} = s(\lambda) |\psi_1\rangle\langle \psi_1| + \frac{1-s(\lambda)}{4} \mathbb{I}_4  \, \, \, \forall i \in \{1,2,3,4\} \nonumber \\
 & \text{with} \, \, s(\lambda)=\frac{1}{2}\left[1-\lambda+\sqrt{(1-\lambda)(1+3\lambda)} \right].
\end{align}
Hence, each of $\rho_{14|i}$, $\rho_{12|i}$ and $\rho_{34|i}$ belongs to the Werner class of states for all $i \in \{1,2,3,4\}$. Another important point to be stressed here is that $\rho_{14|i}$ depends on the outcome $i$, whereas $\rho_{12|i}$ and $\rho_{34|i}$ are independent of $i$. \\

\textbf{Entanglement:}
Next, we evaluate the amount of entanglement of $\rho_{14|i}$, $\rho_{12|i}$ and $\rho_{34|i}$  for all $i \in \{1,2,3,4\}$. From (\ref{eeq}), we get
\begin{equation}
 \label{29}
 \mathcal{E}_{\rho_{14|i}}=\frac{3\lambda-1}{2} \, \, \, \forall i \in \{1,2,3,4\}.
\end{equation}
Each $\rho_{14|i}$ is entangled if and only if $ \frac{1}{3} < \lambda \leq 1$. 
Hence, it is clearly reflected that the state $\rho_{14|i}$ is entangled if and only if Bob's measurement is entangled. 

In the case of $\rho_{12|i}$ or $\rho_{34|i}$, the amount of entanglement is given by,
\begin{align}
 \label{30}
 \mathcal{E}_{\rho_{12|i}}= \mathcal{E}_{\rho_{34|i}}= \frac{3 s(\lambda)-1}{2} \, \, \, \forall i \in \{1,2,3,4\}
\end{align}
with $s(\lambda)$ being defined earlier. From the above Eq.(\ref{30}), it follows that the states  $\rho_{12}$ and $\rho_{34}$ are entangled if and only if $0\leq\lambda<0.91$.

{\centering
   \begin{figure}[t!]
 \includegraphics[width=7cm, height=5cm]{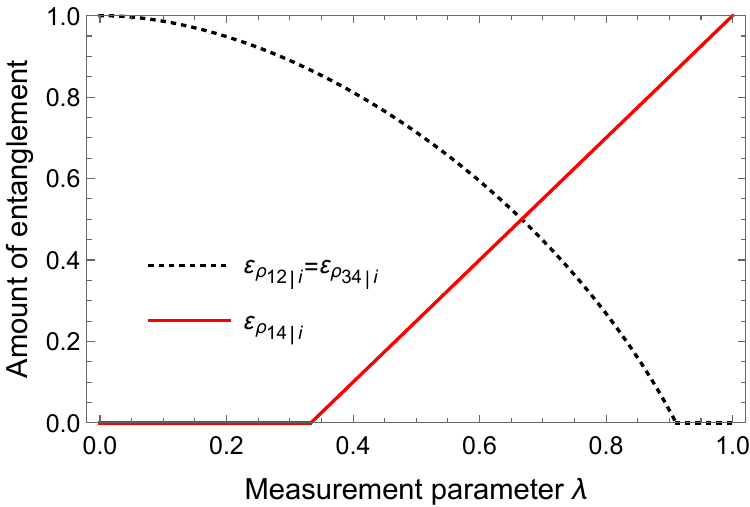}
 \caption[]{Variation of negativity ($\mathcal{E}_{\rho_{14|i}}$, $\mathcal{E}_{\rho_{12|i}}$ and $\mathcal{E}_{\rho_{34|i}}$) with the measurement parameter $\lambda$ for the states  $\rho_{12|i}$, $\rho_{34|i}$ and $\rho_{14|i}$. Here $\rho_{\alpha\beta|i}$ denotes the post-measurement state for the pair ($\alpha,\beta$) after Bob gets the outcome $i$ ($i \in \{1,2,3,4\}$) contingent upon performing the POVM (\ref{wernerbasis}) and communicates the outcome to Alice and Charlie. It is noteworthy that this plot is the same for all possible outcomes $i \in \{1,2,3,4\}$ of Bob's measurement.} 
 \label{fig-negativity}
\end{figure} 
}

We show the variation of $\mathcal{E}_{\rho_{14|i}}$, $\mathcal{E}_{\rho_{12|i}}$ and $\mathcal{E}_{\rho_{34|i}}$ with $\lambda$ in Fig. \ref{fig-negativity}.  From this figure, it is evident that all the resulting states $\rho_{12|i}$, $\rho_{34|i}$ and $\rho_{14|i}$ are simultaneously entangled when $\frac{1}{3}<\lambda<0.91$. In other words, there exists a range of the measurement strength $\lambda$ when the particle of Alice (Charlie) is simulatenously entangled with the particles of Bob and Charlie (Alice)- which is never possible in the usual entanglement swapping scenario with $\lambda=1$.  Moreover, when $\lambda=\frac{2}{3}$, all these three resulting states shared between different parties  have the equal amount of entanglement- $\mathcal{E}_{\rho_{14|i}} = \mathcal{E}_{\rho_{12|i}} = \mathcal{E}_{\rho_{34|i}} =\frac{1}{2}$. These aspects are the same for all possible outcomes  $i \in \{1,2,3,4\}$.




\textbf{EPR Steering:} We now move to analyze the EPR steering of all the post-measurement states which we obtained earlier. For the state $\rho_{14|i}$, the eigenvalues of the matrix $T_{\rho_{14|i}}^TT_{\rho_{14|i}}$ are given by, $t_{1}=\lambda^2$, $t_{2}=\lambda^2$ and $t_{3}=\lambda^2$. Hence, using   Eq.(\ref{ssq}) we have
\begin{equation}
 \label{29}
 \mathcal{S}_{\rho_{14|i}}=\max{\Big\{0,\frac{\sqrt{3}\lambda-1}{\sqrt{3}-1}\Big\}} \, \, \, \forall i \in \{1,2,3,4\}.
\end{equation}
Hence, the state $\rho_{14|i}$ is steerable with respect to the quantum violation of the $3$-setting linear steering inequality \cite{CJWR} when $\frac{1}{\sqrt{3}} < \lambda \leq 1$.

{\centering
\begin{figure}[t!]
 \includegraphics[width=7cm, height=5cm]{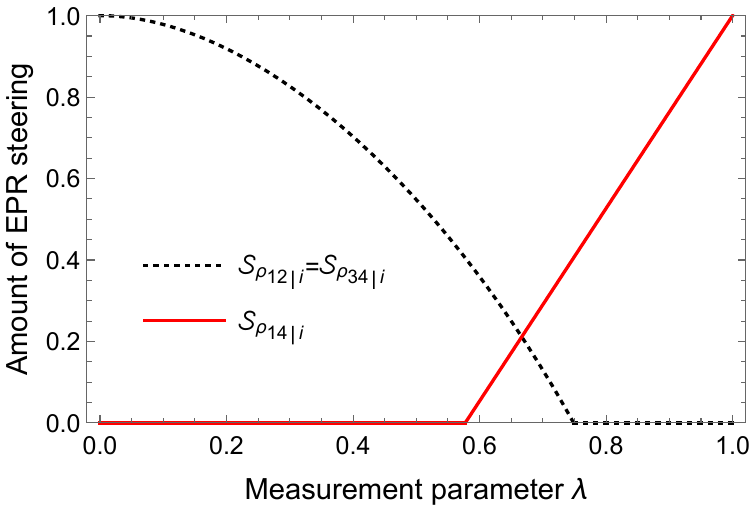}
 \caption[]{Variation of the amount of EPR steering ($\mathcal{S}_{\rho_{14|i}}$, $\mathcal{S}_{\rho_{12|i}}$ and $\mathcal{S}_{\rho_{34|i}}$) with the measurement parameter $\lambda$ for the states states $\rho_{12|i}$, $\rho_{34|i}$ and $\rho_{14|i}$. Here $\rho_{\alpha\beta|i}$ denotes the post-measurement state for the pair ($\alpha,\beta$) after Bob gets the outcome $i$ ($i \in \{1,2,3,4\}$) contingent upon performing the POVM (\ref{wernerbasis}) and communicates the outcome to Alice and Charlie. This plot remains the same for all possible outcomes $i \in \{1,2,3,4\}$ of Bob's measurement.} 
 \label{fig-steering}
\end{figure}
}

Next, let us focus on the states $\rho_{12|i}$ and $\rho_{34|i}$. For each of these two states, the eigenvalues of the matrix $T_{\rho_{m \, m+1|i}}^TT_{\rho_{m \, m+1|i}}$ (with $m \in \{1,3\}$) are given by, $t_{1}=\left[s(\lambda)\right]^2$, $t_{2}=\left[s(\lambda)\right]^2$ and $t_{3}=\left[s(\lambda)\right]^2$, where $s(\lambda)=\frac{1}{2}\left[1-\lambda+\sqrt{(1-\lambda)(1+3\lambda)} \right]$. With these, we have
\begin{equation}
 \label{301}
 \mathcal{S}_{\rho_{12|i}}= \mathcal{S}_{\rho_{34|i}}= \max{\Big\{0,\frac{\sqrt{3} s(\lambda)-1}{\sqrt{3}-1}\Big\}} \, \, \, \forall i \in \{1,2,3,4\}.
\end{equation}
The states $\rho_{12|i}$ and $\rho_{34|i}$ are steerable when $s(\lambda) >  \frac{1}{\sqrt{3}}$, i.e., when $0 \leq \lambda < 0.75$.

We have plotted  the variation of $\mathcal{S}_{\rho_{14|i}}$, $\mathcal{S}_{\rho_{12|i}}$ and $\mathcal{S}_{\rho_{34|i}}$ with $\lambda$ in Fig. \ref{fig-steering}.  From this figure, it can be observed that all the three states $\rho_{12|i}$, $\rho_{34|i}$ and $\rho_{14|i}$ are simultaneously EPR steerable when $\frac{1}{\sqrt{3}}<\lambda<0.75$. For $\lambda=\frac{2}{3}$, all these three resulting states  have the same amount of EPR steerability. Similar to the case of entanglement, here also, these features are the same for all possible outcomes  $i \in \{1,2,3,4\}$. \\


\textbf{Bell nonlocality:} Next, we will focus on Bell-nonlocality of the post-measurement states. Using the eigenvalues of the matrix $T_{\rho_{14|i}}^TT_{\rho_{14|i}}$ derived earlier, we get
\begin{equation}
 \label{29b}
 \mathcal{N}_{\rho_{14|i}}=\max{\Big\{0,\frac{\sqrt{2}\lambda-1}{\sqrt{2}-1}\Big\}} \, \, \, \forall i \in \{1,2,3,4\}.
\end{equation}
Hence, $\rho_{14|i}$ is Bell nonlocal (with respect to the Bell-CHSH inequality) when $\frac{1}{\sqrt{2}} < \lambda \leq 1$.

{\centering
\begin{figure}[t!]
 \includegraphics[width=7cm, height=5cm]{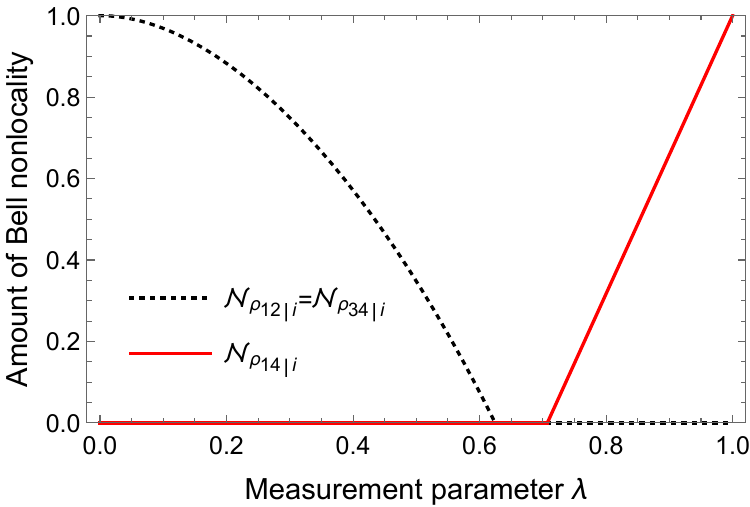}
 \caption[]{Variation of Bell nonlocality ($\mathcal{N}_{\rho_{14|i}}$, $\mathcal{N}_{\rho_{12|i}}$ and $\mathcal{N}_{\rho_{34|i}}$) with the measurement parameter $\lambda$ for the states states $\rho_{12|i}$, $\rho_{34|i}$ and $\rho_{14|i}$. Here $\rho_{\alpha\beta|i}$ denotes the post-measurement state for the pair ($\alpha,\beta$) after Bob gets the outcome $i$ ($i \in \{1,2,3,4\}$) contingent upon performing the POVM (\ref{wernerbasis}) and communicates the outcome to Alice and Charlie. This plot remains the same for all possible outcomes $i \in \{1,2,3,4\}$ of Bob's measurement.} 
 \label{fig-nonlocality}
\end{figure}
}

Similarly, for the states $\rho_{12|i}$ and $\rho_{34|i}$, we have
\begin{equation}
 \label{301b}
 \mathcal{N}_{\rho_{12|i}}= \mathcal{N}_{\rho_{34|i}}= \max{\Big\{0,\frac{\sqrt{2} s(\lambda)-1}{\sqrt{2}-1}\Big\}} \, \, \, \forall i \in \{1,2,3,4\},
\end{equation}
where $s(\lambda)=\frac{1}{2}\left[1-\lambda+\sqrt{(1-\lambda)(1+3\lambda)} \right]$. These two states are Bell nonlocal  when $s(\lambda) >  \frac{1}{\sqrt{2}}$, i.e., when $0 \leq \lambda < 0.62$.

We have plotted  the variation of $\mathcal{N}_{\rho_{14|i}}$, $\mathcal{N}_{\rho_{12|i}}$ and $\mathcal{N}_{\rho_{34|i}}$ with $\lambda$ in Fig. \ref{fig-nonlocality}.  Unlike the case of entanglement and EPR steering, there does not exist any range of $\lambda$, where all the three states $\rho_{12|i}$, $\rho_{34|i}$ and $\rho_{14|i}$ are simultaneously Bell nonlocal with respect to quantum violation of the Bell-CHSH inequality. In particular, when $0.62 \leq \lambda \leq  \frac{1}{\sqrt{2}}$, none of the three states $\rho_{14|i}$, $\rho_{12|i}$ and $\rho_{34|i}$  is Bell nonlocal.


From Table \ref{tab1}, it can be observed that these analyses provide with a tool for constructing different types of quantum correlations in different branches of a network by fine-tuning the measurement strength $\lambda$. For example, if one chooses the POVM (\ref{wernerbasis}) with $\lambda \leq \frac{1}{\sqrt{2}}$, then after completion of the protocol, the pair ($1,4$) becomes entangled without Bell-nonlocality (with respect to the Bell-CHSH inequality). Similarly, if one chooses $\lambda \leq \frac{1}{\sqrt{3}}$, then the pair ($1,4$) is entangled without EPR steering (with respect to the $3$-setting steering inequality) and Bell-nonlocality (with respect to the Bell-CHSH inequality). In a similar way, entanglement without steerability or entanglement without Bell nonlocality can be established between the pair ($1,2$) or ($3,4$) by appropriately choosing $\lambda$. 

\begin{table}[t]
{\centering
 \begin{tabular}{ | m{1.0cm} | m{2.5cm}| m{2.0cm} | m{2.0cm} | } 
 \hline
 States & Entangled & Steerable & Bell-nonlocal\\
  & when & when & when \\
 \hline
 $\rho_{14|i}$ & $\frac{1}{3}<\lambda\leq 1$ & $\frac{1}{\sqrt{3}}<\lambda\leq 1$ & $\frac{1}{\sqrt{2}}<\lambda\leq1$\\ 
 \hline
 $\rho_{12|i}$ & $0\leq\lambda<0.91$ & $0\leq\lambda<0.75$ & $0\leq\lambda<0.62$\\ \hline
 $\rho_{34|i}$ & $0\leq\lambda<0.91$ & $0\leq\lambda<0.75$ & $0\leq\lambda<0.62$\\
 \hline
 \end{tabular}
}
\caption{The ranges of the measurement strength $\lambda$ for which the states $\rho_{12|i}$, $\rho_{34|i}$ and $\rho_{14|i}$ are entangled, or EPR steerable, or Bell-nonlocal for all possible $i \in \{1,2,3,4\}$. Here $\rho_{\alpha\beta|i}$ denotes the post-measurement state for the pair ($\alpha,\beta$) after Bob gets the outcome $i$ ($i \in \{1,2,3,4\}$) contingent upon performing the POVM (\ref{wernerbasis}) and communicates the outcome to Alice and Charlie.}
\label{tab1}
\end{table}

\subsection*{Case II}

We now consider another class of POVM $E\equiv \lbrace E_{1}, E_2, E_3, E_4 \rbrace$ with
\begin{align}
 &E_1=x|\psi_1^{(\lambda)} \rangle \langle \psi_1^{(\lambda)}| + \frac{y_1(1-x)}{y_1+y_2} |\psi_2^{(\lambda)} \rangle\langle\psi_2^{(\lambda)}|  + \frac{y_2(1-x)}{y_1+y_2} |\phi_3\rangle\langle\phi_3|, \nonumber\\
 &E_2=x|\psi_2^{(\lambda)} \rangle\langle\psi_2^{(\lambda)} | +\frac{y_1(1-x)}{y_1+y_2} |\psi_1^{(\lambda)} \rangle \langle \psi_1^{(\lambda)}|   + \frac{y_2(1-x)}{y_1+y_2} |\phi_4\rangle\langle\phi_4|, \nonumber\\
 &E_3=x|\psi_3^{(\lambda)} \rangle \langle \psi_3^{(\lambda)}| + \frac{y_1(1-x)}{y_1+y_2} |\psi_4^{(\lambda)}\rangle \langle\psi_4^{(\lambda)}|   + \frac{y_2(1-x)}{y_1+y_2} |\phi_1\rangle\langle\phi_1|, \nonumber\\
 &E_4=x|\psi_4^{(\lambda)} \rangle\langle\psi_4^{(\lambda)}| + \frac{y_1(1-x)}{y_1+y_2}  |\psi_3^{(\lambda)}\rangle\langle\psi_3^{(\lambda)}|   +  \frac{y_2(1-x)}{y_1+y_2} |\phi_2\rangle\langle\phi_2|, 
 \label{povm2}
\end{align}
where $x=0.3$, $y_1=\frac{2+2\sqrt{1-\lambda}-\lambda}{4}$, $y_2=\frac{\lambda}{4}$ with $0\leq\lambda\leq 1$. We choose  $|\psi_i^{(\lambda)}\rangle$ from the set $\{|\psi_i^{(\lambda)}\rangle, i=1,...4\}$ and $|\phi_i\rangle$ from the set $\{|\phi_i\rangle, i=1,...4\}$ that are given by,
\begin{equation}
 \begin{aligned}[t]
  |\psi_1^{(\lambda)}\rangle&=a|00\rangle-b|11\rangle;\\
  |\psi_2^{(\lambda)}\rangle&=b|00\rangle+a|11\rangle;
 \end{aligned}
  \qquad 
  \begin{aligned}[t]
   |\psi_3^{(\lambda)}\rangle&=a|01\rangle-b|10\rangle;\\
  |\psi_4^{(\lambda)}\rangle&=b|01\rangle+a|10\rangle;
  \end{aligned} \nonumber 
 \end{equation}
 with 
 \begin{equation}
  a=\frac{\sqrt{1-\sqrt{1-\lambda}}}{\sqrt{2}}, \hspace{1.3cm}  b=\frac{\sqrt{1+\sqrt{1-\lambda}}}{\sqrt{2}},  \nonumber 
 \end{equation}
 and 
\begin{equation}
 \begin{aligned}[t]
  |\phi_1\rangle&=|00\rangle;\\
  |\phi_2\rangle&=|11\rangle;
 \end{aligned}
  \qquad 
  \begin{aligned}[t]
   |\phi_3\rangle&=|01\rangle;\\
  |\phi_4\rangle&=|10\rangle.
  \end{aligned} \nonumber 
 \end{equation}
 The above POVM is characterized by the single parameter $\lambda$.
 
 In this case, it can be shown that the probability of getting the $i$th outcome is $p_i= \frac{1}{4}$ for all $i \in \{1,2,3,4\}$.

Next, we will find out the post-measurement states $\rho_{12|i}$, $\rho_{34|i}$ and $\rho_{14|i}$ after Bob performs the above POVM on the pair ($2,3$) and communicates the outcome to Alice, Charlie. From Eq.(\ref{post14}), we get
\begin{align}
 &\rho_{14|1}=x|\psi_1^{(\lambda)} \rangle \langle \psi_1^{(\lambda)}| + \frac{y_1(1-x)}{y_1+y_2} |\psi_2^{(\lambda)} \rangle\langle\psi_2^{(\lambda)}| + \frac{y_2(1-x)}{y_1+y_2} |\phi_3\rangle\langle\phi_3|, \nonumber \\
 &\rho_{14|2}=x|\psi_2^{(\lambda)} \rangle\langle\psi_2^{(\lambda)} | +\frac{y_1(1-x)}{y_1+y_2} |\psi_1^{(\lambda)} \rangle \langle \psi_1^{(\lambda)}|   + \frac{y_2(1-x)}{y_1+y_2} |\phi_4\rangle\langle\phi_4|, \nonumber\\
 &\rho_{14|3}=x|\psi_3^{(\lambda)} \rangle \langle \psi_3^{(\lambda)}| + \frac{y_1(1-x)}{y_1+y_2} |\psi_4^{(\lambda)}\rangle \langle\psi_4^{(\lambda)}|   + \frac{y_2(1-x)}{y_1+y_2} |\phi_1\rangle\langle\phi_1|, \nonumber\\
 &\rho_{14|4}=x|\psi_4^{(\lambda)} \rangle\langle\psi_4^{(\lambda)}| + \frac{y_1(1-x)}{y_1+y_2}  |\psi_3^{(\lambda)}\rangle\langle\psi_3^{(\lambda)}|   +  \frac{y_2(1-x)}{y_1+y_2} |\phi_2\rangle\langle\phi_2|,
 \end{align}
 
 From Eqs.(\ref{post12}) and (\ref{post34}), we get 
 \begin{align}
 \label{eq21}
 \rho_{12|1}&= (e^2+f^2)|\xi_1\rangle\langle\xi_1|+h^2|\phi_3\rangle\langle\phi_3|+h^2|\phi_4\rangle\langle\phi_4|+g^2|\phi_1\rangle\langle\phi_1|, \nonumber\\
 \rho_{34|1}&= (e^2+g^2)|\xi_2\rangle\langle\xi_2|+h^2|\phi_3\rangle\langle\phi_3|+h^2|\phi_4\rangle\langle\phi_4|+f^2|\phi_2\rangle\langle\phi_2|, \nonumber\\
&\text{with} \, \, \, |\xi_1\rangle=\frac{e}{\sqrt{e^2+f^2}}|00\rangle+\frac{f}{\sqrt{e^2+f^2}}|11\rangle , \nonumber \\
 & \hspace{0.75cm} |\xi_2\rangle=\frac{g}{\sqrt{e^2+g^2}}|00\rangle+\frac{e}{\sqrt{e^2+g^2}}|11\rangle,  \nonumber \\
 & \hspace{0.7cm} e=\sqrt{\frac{y_2(1-x)}{y_1+y_2}},  \nonumber \\
 & \hspace{0.7cm} f= a^2 \sqrt{\frac{y_1(1-x)}{y_1+y_2}} +b^2\sqrt{x},  \nonumber \\
 & \hspace{0.7cm} g= b^2 \sqrt{\frac{y_1(1-x)}{y_1+y_2}} +a^2\sqrt{x}, \nonumber \\
 & \hspace{0.7cm} h= ab \sqrt{\frac{y_1(1-x)}{y_1+y_2}} -ab \sqrt{x}.
 \end{align}
 Similarly, $\rho_{12|i}$, $\rho_{34|i}$ for other values of $i$ can be evaluated. Unlike the previous POVM, here the states $\rho_{12|i}$ and $\rho_{34|i}$ are different as the POVM (\ref{povm2}) is asymmetric with respect to the two particles of the pair ($2,3$). This is due to the fact that the set $\{|\phi_i\rangle, i=1,...4\}$ is not symmetric. Note that each of $\rho_{12|i}$, $\rho_{34|i}$ and $\rho_{14|i}$ is mixed state for $0 \leq \lambda \leq 1$.


\textbf{Entanglement:} Let us first focus on the entanglement of the states $\rho_{12|i}$, $\rho_{34|i}$ and $\rho_{14|i}$. The amount of entanglement of the state $\rho_{14|i}$ is given by,
\begin{align}
& \mathcal{E}_{\rho_{14|i}}= \sqrt{4 q^2+r^2}-r \nonumber \\
 & \text{with} \, \, q = \frac{ab [y_1(1-2x)- x y_2]}{y_1+y_2}, \, \, r= \frac{y_2(1-x)}{y_1+y_2} \nonumber \\
 & \hspace{4.5cm} \, \, \, \forall i \in \{1,2,3,4\}.
 \label{ent214}
\end{align}

Next, the amount of entanglement of the states $\rho_{12|i}$ and $\rho_{34|i}$ are given by,
\begin{equation}
 \mathcal{E}_{\rho_{12|i}}=2(ef-h^2) \, \, \, \forall i \in \{1,2,3,4\},
 \label{ent212}
\end{equation}
and 
\begin{equation}
 \mathcal{E}_{\rho_{34|i}}=2(eg-h^2)\, \, \, \forall i \in \{1,2,3,4\}
\label{ent234}
\end{equation}
respectively, where $e$, $f$, $g$, $h$ are defined earlier. Hence, though the structures of each of the states $\rho_{12|i}$, $\rho_{34|i}$ and $\rho_{14|i}$ is different for different $i$, each of $\mathcal{E}_{\rho_{14|i}}$, $\mathcal{E}_{\rho_{12|i}}$ and $\mathcal{E}_{\rho_{34|i}}$ does not depend on $i$. 

{\centering
\begin{figure}[t!]
 \includegraphics[width=7cm, height=5cm]{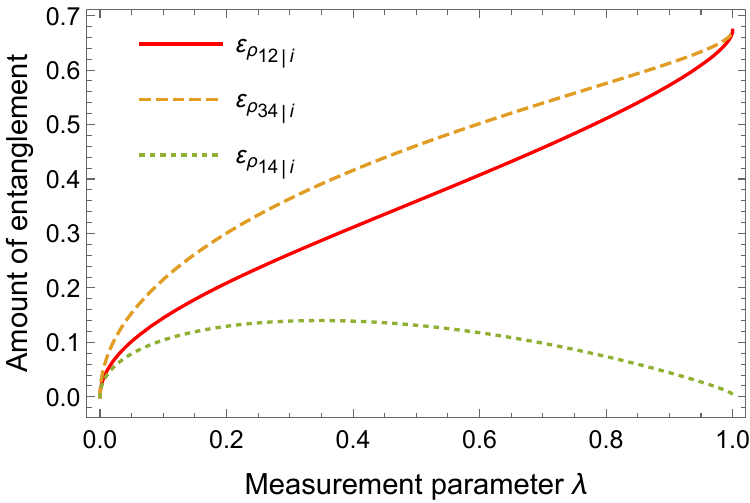}
 \label{fig-entangle1}
 \caption[]{Variation of negativity ($\mathcal{E}_{\rho_{14|i}}$, $\mathcal{E}_{\rho_{12|i}}$ and $\mathcal{E}_{\rho_{34|i}}$) with the measurement parameter $\lambda$ for the states  $\rho_{12|i}$, $\rho_{34|i}$ and $\rho_{14|i}$. Here $\rho_{\alpha\beta|i}$ denotes the post-measurement state for the pair ($\alpha,\beta$) after Bob gets the outcome $i$ ($i \in \{1,2,3,4\}$) contingent upon performing the POVM (\ref{povm2}) with $x=0.3$ and communicates the outcome to Alice and Charlie. This plot is the same for all possible outcomes $i \in \{1,2,3,4\}$ of Bob's measurement.} 
 \label{fig-entangle1}
\end{figure}
}

Putting $x=0.3$, we have plotted the variation of $\mathcal{E}_{\rho_{14|i}}$, $\mathcal{E}_{\rho_{14|i}}$ and $\mathcal{E}_{\rho_{14|i}}$ with $\lambda$ in Fig. \ref{fig-entangle1}. It can be observed that each of the above three states are entangled for the whole range of $\lambda$ except at $\lambda=0$. 

This result demonstrates the following interesting aspect of the generalized entanglement swapping protocol with the POVM (\ref{povm2}). This protocol creates entanglement in the pair ($1,4$). However, it never completely destroys the entanglement content in the pairs ($1,2$) and ($3,4$) for whole range of the measurement parameter $\lambda$ except at $\lambda=0$. 


\textbf{EPR steering:} Next, we will evaluate EPR steerability of different post-measurement states. It can be checked that the eigenvalues of the matrix $T_{\rho_{14|i}}^TT_{\rho_{14|i}}$ for the state $\rho_{14|i}$ are given by,
\begin{eqnarray}
\label{eq20}
t_1&=&\Bigg(\frac{2ab\big[y_2 x+y_1(2x-1)\big]}{y_1+y_2}\Bigg)^2,\nonumber\\
t_2&=&\Bigg(\frac{2ab\big[y_2 x+y_1(2x-1)\big]}{y_1+y_2}\Bigg)^2,\nonumber\\
t_3&=&\Bigg(\frac{y_1+y_2(2x-1)}{y_1+y_2}\Bigg)^2 \, \, \, \forall i \in \{1,2,3,4\}.
\label{t14}
\end{eqnarray}
The eigenvalues of the matrix $T_{\rho_{12|i}}^TT_{\rho_{12|i}}$ for the state $\rho_{12|i}$ are given by,
\begin{eqnarray}
\label{eq22}
t_1&=&4e^2f^2\nonumber\\
t_2&=&4e^2f^2\nonumber\\
t_3&=&(e^2+f^2+g^2-2h^2)^2  \, \, \, \forall i \in \{1,2,3,4\}.
\label{t12}
\end{eqnarray}
And finally, the eigenvalues of the matrix $T_{\rho_{34|i}}^TT_{\rho_{34|i}}$ for the state $\rho_{34|i}$ are given by,
\begin{eqnarray}
\label{eq24}
t_1&=&4e^2g^2\nonumber\\
t_2&=&4e^2g^2\nonumber\\
t_3&=&(e^2+f^2+g^2-2h^2)^2  \, \, \, \forall i \in \{1,2,3,4\}.
\label{t34}
\end{eqnarray}
Using the above expressions,  $\mathcal{S}_{\rho_{14|i}}$, $\mathcal{S}_{\rho_{12|i}}$ and $\mathcal{S}_{\rho_{34|i}}$ can be evaluated by putting $x=0.3$. 
Interestingly, in this case, it can be shown that $\mathcal{S}_{\rho_{14|i}}= 0$  for the whole range of $\lambda \in [0,1]$. On the other hand, $\mathcal{S}_{\rho_{12|i}}>0$ and $\mathcal{S}_{\rho_{34|i}} > 0$ for the whole range of $\lambda$ except at $\lambda=0$.  \\

\textbf{Bell nonlocality:} Using Eqs.(\ref{t14}), (\ref{t12}) and (\ref{t34}) and putting $x=0.3$, it can be checked that each of the two states $\rho_{12|i}$ and $\rho_{34|i}$ is Bell nonlocal for the whole range of $\lambda$ except at $\lambda=0$. On the other hand, the state $\rho_{14|i}$ is not Bell nonlocal for the whole range of $\lambda$. In other words,  $\mathcal{N}_{\rho_{14|i}}=0$ for all $\lambda \in [0,1]$; and $\mathcal{N}_{\rho_{12|i}}>0$ and $\mathcal{N}_{\rho_{34|i}} > 0$ for all $\lambda \in (0,1]$. 

\begin{table}[t]
{\centering
 \begin{tabular}{ | m{1.0cm} | m{1.8cm}| m{2.0cm} | m{2.0cm} | } 
 \hline
 States & Entangled & Steerable & Bell-nonlocal\\
  & when & when & when \\
 \hline
 $\rho_{14|i}$ & $0<\lambda\leq1$ & Never & Never\\ 
 \hline
 $\rho_{12|i}$ & $0<\lambda\leq1$ & $0<\lambda\leq1$ & $0<\lambda\leq1$\\ 
 \hline$\rho_{34|i}$ & $0<\lambda\leq1$ & $0<\lambda\leq1$ & $0<\lambda\leq1$\\ 
 \hline
 \end{tabular}
}
 \caption{The ranges of the measurement parameter $\lambda$ for which the states $\rho_{12|i}$, $\rho_{34|i}$ and $\rho_{14|i}$ are entangled, or EPR steerable, or Bell-nonlocal for all possible $i \in \{1,2,3,4\}$. Here $\rho_{\alpha\beta|i}$ denotes the post-measurement state for the pair ($\alpha,\beta$) after Bob gets the outcome $i$ ($i \in \{1,2,3,4\}$) contingent upon performing the POVM (\ref{povm2})  with $x=0.3$ and communicates the outcome to Alice and Charlie.}
 \label{tab2}
\end{table}

Therefore, the above results indicate that the resulting state $\rho_{14|i}$ (for all possible outcomes $i$) after completion of our generalized entanglement swapping protocol contingent upon using the POVM (\ref{povm2}) with $x=0.3$ is entangled, but does not violate the $3$-setting steering inequality \cite{CJWR} or the Bell-CHSH inequality. This is summarized in Table \ref{tab2}. Interestingly, this holds for the whole range of $\lambda$ except at $\lambda=0$.  Hence, this process leads to the generation of a single parameter ($\lambda$) family of mixed entangled states that does not violate the $3$-setting linear steering inequality or the Bell-CHSH inequality for the whole range of the state parameter, except for case when the state parameter is equal to zero. This class of mixed entangled states has another interesting feature- entanglement increases with an increase in the state parameter $\lambda$ whenever $\lambda \leq 0.34$. However, when $\lambda > 0.34$, then entanglement decreases with an increase in $\lambda$. Hence, it is implies that this class of states never violates the $3$-setting linear steering inequality or the Bell-CHSH inequality even when the entanglement content is varied (by varying $\lambda$).

Similarly, from Table \ref{tab2}, we see that the resulting states $\rho_{12|i}$ and $\rho_{34|i}$ (for all possible outcomes $i$) are Bell nonlocal for the whole range of $\lambda$ except $\lambda=0$. Thus, this generalized entanglement swapping protocol with the POVM (\ref{povm2}) also provides with a tool to generate a single parameter ($\lambda$) family of mixed entangled states that is Bell nonlocal for the whole range of the state parameter, except for $\lambda=0$.

Finally, we would like to point out that this generalized entanglement swapping protocol generates entanglement in the pair ($1,4$), but never generates EPR steering (with respect to quantum violation of the $3$-setting linear steering inequality) or Bell nonlocality (with respect to quantum violation of the Bell-CHSH inequality) in ($1,4$). On the other hand, this protocol never completely destroys Bell nonlocality in each of the pairs ($1,2$) and ($3,4$). This is valid for the whole range of the measurement parameter $\lambda$, except at $\lambda = 0$.

{\centering
\begin{figure}[t!]
 \includegraphics[width=7cm, height=5cm]{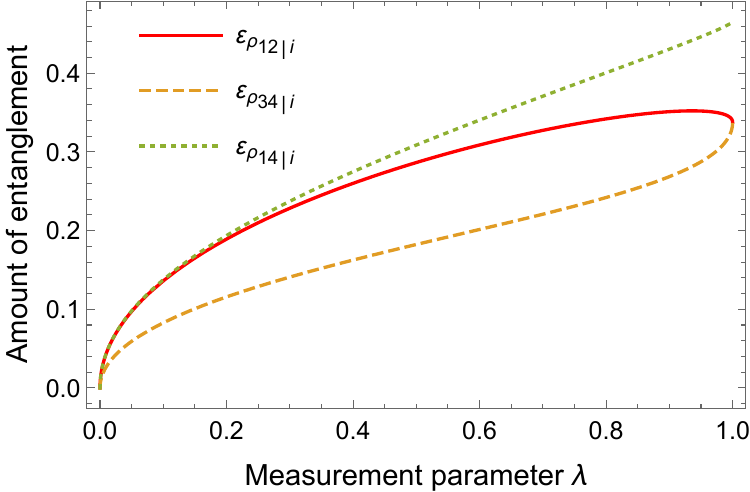}
 \label{fig-entangle2}
 \caption[]{Variation of negativity ($\mathcal{E}_{\rho_{14|i}}$, $\mathcal{E}_{\rho_{12|i}}$ and $\mathcal{E}_{\rho_{34|i}}$) with the measurement parameter $\lambda$ for the states  $\rho_{12|i}$, $\rho_{34|i}$ and $\rho_{14|i}$. Here $\rho_{\alpha\beta|i}$ denotes the post-measurement state for the pair ($\alpha,\beta$) after Bob gets the outcome $i$ ($i \in \{1,2,3,4\}$) contingent upon performing the POVM (\ref{povm2}) with $x=0.725$ and communicates the outcome to Alice and Charlie. This plot is the same for all possible outcomes $i \in \{1,2,3,4\}$ of Bob's measurement.}  
 \label{figent3}
\end{figure}
}

\subsection*{Case III}

Now, we consider  the POVM mentioned in (\ref{povm2}) with $x=0.725$. For this POVM, we will now determine how different quantum correlations are created or destroyed in various pairs of particles. \\

\textbf{Entanglement:} Entanglement of each of the states $\rho_{12|i}$, $\rho_{34|i}$ and $\rho_{14|i}$ can be determined by putting $x = 0.725$ in Eqs.(\ref{ent214}), (\ref{ent212}), (\ref{ent234}). In this case, we get that $\mathcal{E}_{\rho_{14|i}} > 0$, $\mathcal{E}_{\rho_{12|i}}>0$, $\mathcal{E}_{\rho_{34|i}} > 0$ for the whole range of $\lambda$ except at $\lambda=0$. This is valid for any $i \in \{1,2,3,4\}$. The variation of $\mathcal{E}_{\rho_{14|i}}$, $\mathcal{E}_{\rho_{12|i}}$, $\mathcal{E}_{\rho_{34|i}}$ with $\lambda$ is depicted in Fig. \ref{figent3}.

Hence, similar to the previous case, this protocol creates
entanglement in the pair (1, 4) without completely destroying
the entanglement content in the pairs (1, 2) and (3, 4). This feature persists for the whole range of the measurement parameter $\lambda$ except at $\lambda=0$.  

\textbf{EPR steering:} Using the expressions of the eigenvalues of the matrices $T_{\rho_{14|i}}^TT_{\rho_{14|i}}$, $T_{\rho_{12|i}}^TT_{\rho_{12|i}}$ and $T_{\rho_{34|i}}^TT_{\rho_{34|i}}$ mentioned in Eqs.(\ref{t14}), (\ref{t12}), (\ref{t34}) and putting $x=0.725$, we get the following results for all $i \in \{1,2,3,4\}$: $\mathcal{S}_{\rho_{14|i}} > 0$ and $\mathcal{S}_{\rho_{12|i}}>0$ for all $\lambda \in (0,1]$; whereas $\mathcal{S}_{\rho_{34|i}} = 0$ for all $\lambda \in [0,1]$.

\textbf{Bell nonlocality:} Similarly, using Eqs.(\ref{t14}), (\ref{t12}), (\ref{t34}) and putting $x=0.725$, we get the following results in the context of Bell nonlocality for all $i \in \{1,2,3,4\}$: $\mathcal{N}_{\rho_{14|i}} = 0$, $\mathcal{N}_{\rho_{12|i}}=0$ and $\mathcal{N}_{\rho_{34|i}} = 0$ for all $\lambda \in [0,1]$.

\begin{table}[t]
{\centering
 \begin{tabular}{ | m{1.0cm} | m{1.8cm}| m{2.0cm} | m{2.0cm} | } 
 \hline
 States & Entangled & Steerable & Bell-nonlocal\\
  & when & when & when \\
 \hline
 $\rho_{14|i}$ & $0<\lambda\leq1$ & $0<\lambda\leq1$ & Never\\ 
 \hline
 $\rho_{12|i}$ & $0<\lambda\leq1$ & $0<\lambda\leq1$ & Never\\ 
 \hline
 $\rho_{34|i}$ & $0<\lambda\leq1$ & Never & Never\\ 
 \hline
 \end{tabular}
}
 \caption{The ranges of the measurement parameter $\lambda$ for which the states $\rho_{12|i}$, $\rho_{34|i}$ and $\rho_{14|i}$ are entangled, or EPR steerable, or Bell-nonlocal for all possible $i \in \{1,2,3,4\}$. Here $\rho_{\alpha\beta|i}$ denotes the post-measurement state for the pair ($\alpha,\beta$) after Bob gets the outcome $i$ ($i \in \{1,2,3,4\}$) contingent upon performing  the POVM (\ref{povm2})  with $x=0.725$ and communicates the outcome to Alice and Charlie.}
 \label{tab3}
\end{table}

All the above results are summarized in Table \ref{tab3}. This table points out that each of the two resulting state $\rho_{14|i}$ and $\rho_{12|i}$ (for all possible outcomes $i$) after completion of our generalized entanglement swapping protocol contingent upon using the POVM (\ref{povm2}) with $x=0.725$ is entangled as well as steerable, but does not violate the Bell-CHSH inequality.  This holds for the whole range of $\lambda$ except at $\lambda=0$.  Hence, this protocol can be used as a tool to generate  a single parameter  family of EPR steerable mixed states (where the state parameter is $\lambda$) that does not violate the Bell-CHSH inequality for the whole range of the state parameter, except for case when the state parameter is equal to zero. Further,  this class of states never violates the  Bell-CHSH inequality even when the amount of entanglement is varied by varying the state parameter $\lambda$.

On the other hand, the resulting state  $\rho_{34|i}$ (for all possible outcomes $i$) is entangled, but does not violate the $3$-setting steering inequality \cite{CJWR} or the Bell-CHSH inequality even when the entanglement of $\rho_{34|i}$ is varied by varying $\lambda$.

Finally, it should be noted that this generalized entanglement swapping protocol generates entanglement as well as EPR steering in the pair ($1,4$), but never generates Bell nonlocality  in ($1,4$). This protocol never destroys EPR steering in ($1,2$), but completely destroys Bell nonlocality in ($1,2$).  Also, this protocol never destroys entanglement in ($3,4$), but completely destroys EPR steering in ($3,4$). This is valid for the whole range of the measurement parameter $\lambda$, except at $\lambda = 0$. Here, destroying Bell nonlocality and EPR steering is probed with respect to the quantum violations of the Bell-CHSH inequality and the $3$-setting linear steering inequality respectively.

\subsection*{Case IV}

Next, we again consider  the POVM mentioned in (\ref{povm2}), but with $x=0.8$. For this POVM, let us now find out how different quantum correlations are created or destroyed in various pairs in the generalized entanglement swapping scheme. 

\textbf{Entanglement:} Entanglement of each of the states $\rho_{12|i}$, $\rho_{34|i}$ and $\rho_{14|i}$ is determined by taking $x = 0.8$ in Eqs.(\ref{ent214}), (\ref{ent212}), (\ref{ent234}). In this case also, we get that $\mathcal{E}_{\rho_{14|i}} > 0$, $\mathcal{E}_{\rho_{12|i}}>0$, $\mathcal{E}_{\rho_{34|i}} > 0$ for the whole range of $\lambda$ except at $\lambda=0$. This holds for any $i \in \{1,2,3,4\}$. The variation of $\mathcal{E}_{\rho_{14|i}}$, $\mathcal{E}_{\rho_{12|i}}$, $\mathcal{E}_{\rho_{34|i}}$ with $\lambda$ is depicted in Fig. \ref{figent4}.

{\centering
\begin{figure}[t!]
 \includegraphics[width=7cm, height=5cm]{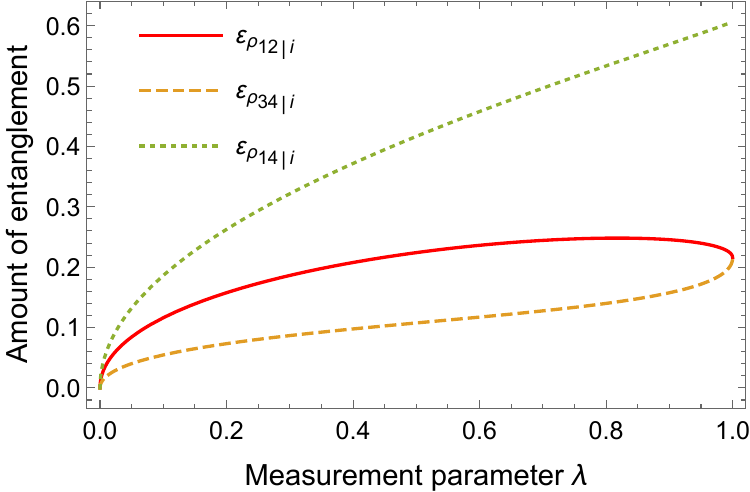}
 \label{fig-entangle2}
 \caption[]{Variation of negativity ($\mathcal{E}_{\rho_{14|i}}$, $\mathcal{E}_{\rho_{12|i}}$ and $\mathcal{E}_{\rho_{34|i}}$) with the measurement parameter $\lambda$ for the states  $\rho_{12|i}$, $\rho_{34|i}$ and $\rho_{14|i}$. Here $\rho_{\alpha\beta|i}$ denotes the post-measurement state for the pair ($\alpha,\beta$) after Bob gets the outcome $i$ ($i \in \{1,2,3,4\}$) contingent upon performing the POVM (\ref{povm2}) with $x=0.8$ and communicates the outcome to Alice and Charlie. This plot is the same for all possible outcomes $i \in \{1,2,3,4\}$ of Bob's measurement.}  
 \label{figent4}
\end{figure}
}

 Hence, in this protocol also,  entanglement is created in the pair ($1,4$), but the entanglement content is not destroyed completely in the pairs ($1,2$) and ($3,4$). This is valid for the whole range of the measurement parameter $\lambda$ except at $\lambda=0$.  

\textbf{EPR steering:} Using the expressions of the eigenvalues of the matrices $T_{\rho_{14|i}}^TT_{\rho_{14|i}}$, $T_{\rho_{12|i}}^TT_{\rho_{12|i}}$ and $T_{\rho_{34|i}}^TT_{\rho_{34|i}}$ mentioned earlier and putting $x=0.8$, we get the following results for all $i \in \{1,2,3,4\}$: $\mathcal{S}_{\rho_{14|i}} > 0$ for all $\lambda \in (0,1]$; whereas $\mathcal{S}_{\rho_{34|i}} = \mathcal{S}_{\rho_{12|i}} = 0$ for all $\lambda \in [0,1]$.

\textbf{Bell nonlocality:} Using Eqs.(\ref{t14}), (\ref{t12}), (\ref{t34}) and putting $x=0.8$, we get the following results in the context of Bell nonlocality for all $i \in \{1,2,3,4\}$: $\mathcal{N}_{\rho_{14|i}} > 0$ for all $\lambda \in (0,1]$, $\mathcal{N}_{\rho_{12|i}}=0$ and $\mathcal{N}_{\rho_{34|i}} = 0$ for all $\lambda \in [0,1]$.

The above results are summarized in Table \ref{tab4}. It is evident from Table \ref{tab4} that  the resulting state $\rho_{14|i}$ (for all possible outcomes $i$), generated after completion of our  protocol using the POVM (\ref{povm2}) with $x=0.8$, is Bell nonlocal for the whole range of $\lambda$ except at $\lambda=0$.  Thus this protocol is an operational tool to   construct   a single parameter  family of  mixed states (where the state parameter is $\lambda$) that is Bell nonlocal for the whole range of the state parameter, except for case when the state parameter is equal to zero. 

On the other hand, each of the states  $\rho_{12|i}$ and $\rho_{34|i}$ (for all possible outcomes $i$) is entangled, but does not violate the $3$-setting steering inequality \cite{CJWR} or the Bell-CHSH inequality even if one varies the entanglement of it. 

Finally, this generalized entanglement swapping protocol always generates Bell nonlocality (EPR steering) in ($1,4$) by completely destroying the Bell nonlocality (EPR steering) in ($1,2$) and ($3,4$) for whole range of the measurement strength $\lambda$. On the other hand, entanglement in ($1,4$) is generated without completely destroying entanglement in ($1,2$) or ($3,4$). Importantly, this holds for the whole range of the measurement parameter $\lambda$, except at $\lambda = 0$. Here also, destroying Bell nonlocality and EPR steering is probed with respect to the quantum violations of the Bell-CHSH inequality and the $3$-setting linear steering inequality respectively.

\begin{table}[t]
{\centering
 \begin{tabular}{ | m{1.0cm} | m{1.8cm}| m{2.0cm} | m{2.0cm} | } 
 \hline
 States & Entangled & Steerable & Bell-nonlocal\\
  & when & when & when \\
 \hline
 $\rho_{14|i}$ & $0<\lambda\leq1$ & $0<\lambda\leq1$ & $0<\lambda\leq1$\\ 
 \hline
 $\rho_{12|i}$ & $0<\lambda\leq1$ & Never & Never\\ 
 \hline
 $\rho_{34|i}$ & $0<\lambda\leq1$ & Never & Never\\ 
 \hline
 \end{tabular}
}
 \caption{The ranges of the measurement parameter $\lambda$ for which the states $\rho_{12|i}$, $\rho_{34|i}$ and $\rho_{14|i}$ are entangled, or EPR steerable, or Bell-nonlocal for all possible $i \in \{1,2,3,4\}$. Here $\rho_{\alpha\beta|i}$ denotes the post-measurement state for the pair ($\alpha,\beta$) after Bob gets the outcome $i$ ($i \in \{1,2,3,4\}$) contingent upon performing the POVM (\ref{povm2})  with $x=0.8$ and communicates the outcome to Alice and Charlie.}
 \label{tab4}
\end{table}


\subsection*{When the initial states are mixed ones}

So far we have studied quantum correlation transfer in the generalized entanglement swapping scenario when the initial states are maximally entangled ones. However, in realistic situations, such maximally entangled states are difficult to be prepared. That is why, we next consider mixed initial states in order to incorporate experimental non-idealness. In what follows, we will show that similar types of quantum correlation transfer is possible in the generalized entanglement swapping scenario involving mixed initial states shared by the pairs (1,2) and (3,4) when the previously mentioned  POVM given by Eq.(\ref{povm2}) is performed on the pair of particles (2,3).

Here, instead of Bell states,  Alice shares a  mixed state $\rho_{12}=m|\phi^+\rangle\langle\phi^+|+(1-m)|\phi^-\rangle\langle\phi^{-}|$ with Bob and  Bob shares another copy of the same state, $\rho_{34}=m|\phi^+\rangle\langle\phi^+|+(1-m)|\phi^-\rangle\langle\phi^{-}|$ with Charlie where $\frac{1}{2}\leq m\leq1$ and $|\phi^{\pm}\rangle=\frac{1}{\sqrt{2}}(|00\rangle\pm|11\rangle)$.

In this scenario, the post-measurement state, when Bob gets the outcome $1$ (associated with $E_1$ mentioned in Eq.(\ref{povm2})) and communicates the measurement outcome to Alice and Charlie,  is given by,
\begin{align}
    \label{eq38}
    \rho_{14|1}=&\frac{(2m-1)^2+1}{2}(g^2+h^2)|\eta_1\rangle\langle\eta_1|\nonumber\\
    &+\frac{1-(2m-1)^2}{2}(f^2+h^2)|\eta_2\rangle\langle\eta_2|+e^2|\phi_3\rangle\langle\phi_3|\nonumber\\
    &\text{with} \,\,\, 
|\eta_1\rangle=\frac{g}{\sqrt{g^2+h^2}}|00\rangle+\frac{h}{\sqrt{g^2+h^2}}|11\rangle,\nonumber\\
&\hspace{.75cm}|\eta_2\rangle=\frac{h}{\sqrt{f^2+h^2}}|00\rangle-\frac{f}{\sqrt{f^2+h^2}}|11\rangle.
\end{align}

Similarly, the other post-measurement states  shared between Alice-Bob and Bob-Charlie are given by,
\begin{align}
    \label{eq39}
    \rho_{12|1}= & m(e^2+f^2)|\zeta_1\rangle\langle\zeta_1|+(1-m)(e^2+f^2)|\zeta_2\rangle\langle\zeta_2|+g^2|\phi_1\rangle\langle\phi_1|\nonumber\\
    &\hspace{0.75cm}+h^2|\phi_3\rangle\langle\phi_3|+h^2|\phi_4\rangle\langle\phi_4|\nonumber\\
    & \hspace{0.6cm} \text{with} \,\,\, 
|\zeta_1\rangle=\frac{e}{\sqrt{e^2+f^2}}|00\rangle+\frac{f}{\sqrt{e^2+f^2}}|11\rangle,\nonumber\\
&\hspace{1.25cm}|\zeta_2\rangle=\frac{e}{\sqrt{e^2+f^2}}|00\rangle-\frac{f}{\sqrt{e^2+f^2}}|11\rangle.
\end{align}
and 
\begin{align}
    \label{eq40}
    \rho_{34|1}&=m(e^2+g^2)|\chi_1\rangle\langle\chi_1|+(1-m)(e^2+g^2)|\chi_2\rangle\langle\chi_2|+f^2|\phi_2\rangle\langle\phi_2|\nonumber\\
    &\hspace{0.75cm}+h^2|\phi_3\rangle\langle\phi_3|+h^2|\phi_4\rangle\langle\phi_4|\nonumber\\
    &\hspace{0.6cm}\text{with} \,\,\, 
|\chi_1\rangle=\frac{g}{\sqrt{e^2+g^2}}|00\rangle+\frac{e}{\sqrt{e^2+g^2}}|11\rangle,\nonumber\\
&\hspace{1.25cm}|\chi_2\rangle=\frac{g}{\sqrt{e^2+g^2}}|00\rangle-\frac{e}{\sqrt{e^2+g^2}}|11\rangle,
\end{align}
where $e$, $f$, $g$, $h$, $|\phi_1\rangle$, $|\phi_2\rangle$, $|\phi_3\rangle$, $|\phi_4\rangle$ are defined earlier. For a fixed value of $m$ and $x$, all the post-measurement states are single parameter families of mixed states.

Similarly, the post-measurement states $\rho_{14|i}$, $\rho_{12|i}$, and $\rho_{34|i}$ for other values of $i\in\{2,3,4\}$ can be evaluated. 

Furthermore, the probability of getting the $i$-th outcome is $p_i=\frac{1}{4}$ for all $i\in\{1,2,3,4\}$. 

Now, it can be checked that one can generate all types of quantum correlated states (i.e., entangled but not steerable; steerable but not Bell-nonlocal; Bell-nonlocal) that are mentioned in Tables \ref{tab2}, \ref{tab3}, \ref{tab4} in the pairs (1,4),  (1,2), (3,4), when the initial states shared by (1,2) and (3,4) are the above-mentioned mixed states with $0.9\leq m\leq0.97$ and when Bob gets any outcome contingent upon performing the POVM (\ref{povm2}) with $x=0.55$, or $x=0.77$, or $x=0.9$.
We  summarize the results in Table \ref{tab5}. 

\begin{table}
\begin{center}
\begin{tabular}{ |m{1cm}|m{1cm}|m{1.8cm}|m{2cm}|m{1.7cm}|} 
\hline
$x$  & States & Entangled when & Steerable when & Bell-nonlocal when \\
\hline
\multirow{3}{4em}{$x=0.55$} & $\rho_{14|i}$ & $0<\lambda\leq1$ &Never&Never\\ 
& $\rho_{12|i}$ & $0<\lambda\leq1$ &$0<\lambda\leq1$&$0<\lambda\leq1$\\
& $\rho_{34|i}$ & $0<\lambda\leq1$ &$0<\lambda\leq1$&$0<\lambda\leq1$\\
\hline
\multirow{3}{4em}{$x=0.77$} & $\rho_{14|i}$ & $0<\lambda\leq1$ &$0<\lambda\leq1$&Never\\ 
& $\rho_{12|i}$ & $0<\lambda\leq1$ &Never&Never\\
& $\rho_{34|i}$ & $0<\lambda\leq1$ &Never&Never\\
\hline
\multirow{3}{4em}{$x=0.9$} & $\rho_{14|i}$ & $0<\lambda\leq1$ &$0<\lambda\leq1$&$0<\lambda\leq1$\\ 
& $\rho_{12|i}$ & $0<\lambda\leq1$ &Never&Never\\
& $\rho_{34|i}$ & $0<\lambda\leq1$ &Never&Never\\
\hline
\end{tabular}
\end{center}
 \caption{The ranges of the measurement parameter $\lambda$ for which the states $\rho_{12|i}$, $\rho_{34|i}$ and $\rho_{14|i}$ are entangled, or EPR steerable, or Bell-nonlocal for all possible $i \in \{1,2,3,4\}$ when eahc of the initial states shared by the pairs (1,2) and (3,4) is $m|\phi^+\rangle\langle\phi^+|+(1-m)|\phi^-\rangle\langle\phi^-|$ with $0.9\leq m\leq 0.97$. Here $\rho_{\alpha\beta|i}$ denotes the post-measurement state for the pair ($\alpha,\beta$) after Bob gets the outcome $i$ ($i \in \{1,2,3,4\}$) contingent upon performing the POVM (\ref{povm2})  with $x=0.55$, or with $x=0.77$, or with  $x=0.9$.}
 \label{tab5}
\end{table}

\subsection*{Probing Network Nonlocality}

 So far, while studying Bell nonlocality and EPR steering, we have not taken into account of the fact that the sources producing the two initial entangled states are in general independent. Now, in order to consider independence of sources, we will study network nonlocality in the scenario considered by us. Before considering the specific scenario, let us assume that two arbitrary two-qubit states are initially shared by the pairs Alice-Bob and Bob-Charlie. This will lead to a general condition for violating the bilocality inequality (\ref{63}) in the generalized entanglement swapping scenario considered by us. 


Let  Alice and Bob share a two-qubit state 
\begin{equation}
    \label{60}
    \rho_{AB}=\frac{1}{4}\left[\mathbb{I}_2 \otimes \mathbb{I}_2 + \mathbf{r}_{A}^{AB}\cdot \boldsymbol{\sigma}\otimes \mathbb{I}_2 + \mathbb{I}_2 \otimes \mathbf{r}_{B}^{AB} \cdot \boldsymbol{\sigma} + \sum_{i,j=1}^3 t_{ij}^{AB} \sigma_i\otimes\sigma_j \right]
\end{equation}
produced by the source $S_1$. Similarly, Bob and Charlie share a two-qubits  state 
\begin{equation}
    \label{61}
    \rho_{BC}=\frac{1}{4}\left[\mathbb{I}_2 \otimes \mathbb{I}_2 +\mathbf{r}_{B}^{BC}\cdot\boldsymbol{\sigma}\otimes \mathbb{I}_2 + \mathbb{I}_2 \otimes \mathbf{r}_{C}^{BC} \cdot \boldsymbol{\sigma} + \sum_{i,j=1}^3 t_{ij}^{BC} \sigma_i\otimes\sigma_j \right]
\end{equation}
produced by  another source $S_2$. Here, $\mathbf{r}_A^{AB}=(r_{A|1}^{AB},r_{A|2}^{AB},r_{A|3}^{AB})$ and $\mathbf{r_B}^{AB}=(r_{B|1}^{AB},r_{B|2}^{AB},r_{B|3}^{AB})$  represents the Bloch vectors of Alice's reduced state and Bob's  reduced state, respectively, in $\rho_{AB}$; $\boldsymbol{\sigma} = (\sigma_1, \sigma_2, \sigma_3)$ is the vector composed of  Pauli matrices; while $t_{ij}^{AB}$ are the terms correspond to the correlation matrix of the state $\rho_{AB}$ and similarly for the state $\rho_{BC}$.  The correlation matrix of the initial states can be written in polar decomposition form,  $t^{AB}=U^{AB}R^{AB}$ where $U^{AB}$ is a unitary matrix and $R^{AB}=\sqrt{t^{AB\dagger}t^{AB}}$. Let  $\tau_{1}^{AB}\geq\tau_{2}^{AB}\geq\tau_{3}^{AB}$ are three non-negative eigenvalues of $R^{AB}$ and $\tau_{1}^{BC}\geq\tau_{2}^{BC}\geq\tau_{3}^{BC}$ are three non-negative eigenvalues of $R^{BC}$.

Now we find out a criteria for violating the bilocality inequality (\ref{63}) when Alice and Charlie measure single qubit projective measurements but Bob does a POVM measurements on two qubits.


\textbf{1st type POVM:} At first, let us consider that Bob performs POVM mentioned in Eq.(\ref{wernerbasis}). Also, consider that Alice performs measurements of the observables $\boldsymbol{\sigma} \cdot \mathbf{u}$ and $\boldsymbol{\sigma} \cdot \mathbf{u}^{\prime}$ depending on whether Alice's input is $\Tilde{x}=0$ or $\Tilde{x} = 1$ respectively. Similarly, Charlie also performs measurements of the observables $\boldsymbol{\sigma} \cdot \mathbf{v}$ and $\boldsymbol{\sigma} \cdot \mathbf{v}^{\prime}$ depending on whether his input is $\Tilde{z}=0$ or $\Tilde{z} = 1$ respectively. Here, $\mathbf{u} = (u_1, u_2, u_3)$, $\mathbf{u}^{\prime} = (u_1^{\prime}, u_2^{\prime}, u_3^{\prime})$, $\mathbf{v} = (v_1, v_2, v_3)$, $\mathbf{v}^{\prime} = (v_1^{\prime}, v_2^{\prime}, v_3^{\prime})$ are unit vectors. Now, using the relations given by Eqs.(\ref{64}-\ref{etop}), it turns out that  $B_0=E_1+E_2-E_3-E_4=\lambda\sigma_3\otimes\sigma_3$ and $B_1=E_1-E_2+E_3-E_4=\lambda\sigma_1\otimes\sigma_1$ where $0\leq\lambda\leq1$.

One important point to be mentioned here is that the bilocality inequality (\ref{63}) is valid when $|\langle B_0\rangle|\leq1$ and $|\langle B_1\rangle|\leq1$ \cite{brain2012}. In the present case, since we have $B_0=\lambda\sigma_3\otimes\sigma_3$ and $B_1=\lambda\sigma_1\otimes\sigma_1$ with $0\leq\lambda\leq1$, the two conditions- $|\langle B_0\rangle|\leq1$ and $|\langle B_1\rangle|\leq1$ are satisfied. Hence, we can safely use the bilocality inequality (\ref{63}).  

Now to get the maximum value of $\mathcal{B}_{bilocal}$ in the present context, we will follow the method mentioned in \cite{gisin2017}. At first, let us calculate the quantity $I$ as follows,
\begin{eqnarray}
\label{67}
I&=&\mbox{Tr}\left[ \left\{ \left(\mathbf{u} + \mathbf{u}^{\prime} \right) \cdot \boldsymbol{\sigma} \otimes \lambda(\sigma_3 \otimes \sigma_3) \otimes \left(\mathbf{v}+\mathbf{v}^{\prime} \right) \cdot \boldsymbol{\sigma} \right\} \left(\rho_{AB}\otimes\rho_{BC} \right) \right]\nonumber\\
&=& \lambda \, \mbox{Tr}\left[ \left\{ (\mathbf{u}+ \mathbf{u}^{\prime}) \cdot \boldsymbol{\sigma} \otimes \sigma_3 \right\}\rho_{AB} \right] \, \, \mbox{Tr} \left[ \left\{ \sigma_3 \otimes (\mathbf{v}+ \mathbf{v}^{\prime}) \cdot \boldsymbol{\sigma} \right\} \rho_{BC} \right]\nonumber\\
&=& \lambda \, \left(\sum_{i=1,2,3}(u_i+u_i^{\prime}) \,  t_{i3}^{AB} \right) \left( \sum_{k=1,2,3} (v_k+v_k^{\prime}) \, t^{BC}_{3k} \right) 
\end{eqnarray}
Similarly, $J$ can be expressed as 
\begin{eqnarray}
\label{68}
J&=&\mbox{Tr}\left[ \left\{ \left(\mathbf{u} - \mathbf{u}^{\prime} \right) \cdot \boldsymbol{\sigma} \otimes \lambda(\sigma_1 \otimes \sigma_1) \otimes \left(\mathbf{v} - \mathbf{v}^{\prime} \right) \cdot \boldsymbol{\sigma} \right\} \left(\rho_{AB}\otimes\rho_{BC} \right) \right]\nonumber\\
&=& \lambda \, \mbox{Tr}\left[ \left\{ (\mathbf{u} - \mathbf{u}^{\prime}) \cdot \boldsymbol{\sigma} \otimes \sigma_1 \right\}\rho_{AB} \right] \, \, \mbox{Tr} \left[ \left\{ \sigma_1 \otimes (\mathbf{v} - \mathbf{v}^{\prime}) \cdot \boldsymbol{\sigma} \right\} \rho_{BC} \right]\nonumber\\
&=& \lambda \, \left(\sum_{i=1,2,3}(u_i -  u_i^{\prime}) \,  t_{i1}^{AB} \right) \left( \sum_{k=1,2,3} (v_k - v_k^{\prime}) \, t^{BC}_{1k} \right) 
\end{eqnarray}

Let the initial states given by Eqs.(\ref{60}) and (\ref{61}) have been defined in such a way that the $Z$ and $X$ Bloch directions of $\rho_{AB}$ are given by the eigenvectors of the matrix $R^{AB}$ associated with the two largest eigenvalues- $\tau_1^{AB}$ and $\tau_2^{AB}$.  Similarly, the $Z$ and $X$ Bloch directions of $\rho_{BC}$ are also given by the eigenvectors of the matrix $R^{BC}$ associated with the two largest eigenvalues- $\tau_1^{BC}$ and $\tau_2^{BC}$.. 

Next, we want to maximize $\mathcal{B}_{bilocal}$ with respect to  $\mathbf{u}$, $\mathbf{u}^{\prime}$, $\mathbf{v}$, and $\mathbf{v}^{\prime}$. From Eqs.(\ref{67}) and (\ref{68}), it is clear that the unit vectors- $\mathbf{u}$, $\mathbf{u}^{\prime}$, $\mathbf{v}$, and $\mathbf{v}^{\prime}$ should lie within the two dimensional subspace spanned by two eigenvectors associated with the two largest eigenvalues. Hence, let us consider that $\mathbf{u}=(\sin\alpha,0,\cos\alpha)$, $\mathbf{u}^{\prime}=(\sin\alpha^{\prime},0,\cos\alpha^{\prime})$, $\mathbf{v}=(\sin\gamma,0,\cos\gamma)$ and $\mathbf{v}^{\prime}=(\sin\gamma^{\prime},0,\cos\gamma^{\prime})$. 
With these, the maximum value of $\mathcal{B}_{bilocal}$ can be found by solving the following: $\partial_{\alpha}\mathcal{B}_{bilocal}=0$, $\partial_{\alpha^{\prime}}\mathcal{B}_{bilocal}=0$, $\partial_{\gamma}\mathcal{B}_{bilocal}=0$ and $\partial_{\gamma^{\prime}}\mathcal{B}_{bilocal}=0$. After doing the above partial differentiation, we get various solutions. It can be checked that among those solutions, the following set gives the maximum of $\mathcal{B}_{bilocal}$: $\alpha=-\alpha^{\prime}$, $\gamma=-\gamma^{\prime}$ and $\cos\alpha=\cos\gamma=\sqrt{\frac{\lambda\tau_{1}^{AB}\tau_1^{BC}}{\lambda\tau_1^{AB}\tau_{1}^{BC}+\lambda\tau_2^{AB}\tau_2^{BC}}}$. And the maximal value of left-hand side of the bilocality inequality (\ref{63}) is given by,
\begin{equation}
    \label{71}
    \mathcal{B}_{bilocal}^{max}=2\sqrt{\lambda\tau_1^{AB}\tau_{1}^{BC}+\lambda\tau_2^{AB}\tau_2^{BC}}.
\end{equation}
The above relation is valid for any two arbitrary two-qubits states shared by Alice-Bob, Bob-Charlie and when Bob performs the two-qubit POVM given by (\ref{wernerbasis}). Hence, the bilocality inequality (\ref{63}) will be violated if 
\begin{equation}
  \lambda\tau_1^{AB}\tau_{1}^{BC}+\lambda\tau_2^{AB}\tau_2^{BC}>1  .
\end{equation}

In our scenario, we have considered maximally entangled Bell pairs: $\rho_{AB}=\rho_{BC}=|\phi^+\rangle\langle\phi^+|$. Hence, from (\ref{71}), the bilocality inequality becomes
\begin{equation}
    \label{72}
    \mathcal{B}_{bilocal}^{max}=2\sqrt{2\lambda}\leq2
\end{equation}
as $\tau_1^{AB}=\tau_2^{AB}=1$ and $\tau_1^{BC}=\tau_2^{BC}=1$ in the present case. Therefore, the bilocality inequality (\ref{63}) is violated if  $\lambda>1/2$. 

\textbf{2nd type POVM:}\hspace{0.2cm} Now, we want to get a criteria to violate the bilocality inequality when Bob performs the POVM given by Eq.(\ref{povm2}) with different values of $x$.  Here also, Alice performs measurements of the observables $\boldsymbol{\sigma} \cdot \mathbf{u}$ and $\boldsymbol{\sigma} \cdot \mathbf{u}^{\prime}$ depending on whether Alice's input is $\Tilde{x}=0$ or $\Tilde{x} = 1$ respectively. Charlie performs measurements of the observables $\boldsymbol{\sigma} \cdot \mathbf{v}$ and $\boldsymbol{\sigma} \cdot \mathbf{v}^{\prime}$ depending on whether his input is $\Tilde{z}=0$ or $\Tilde{z} = 1$ respectively. Using the relations given by Eqs.(\ref{64}-\ref{etop}), it turns out in the present case that $B_1=E_1-E_2+E_3-E_4=\Gamma_{1} \, \sigma_1\otimes\sigma_1 + \Gamma_2 \,  \sigma_3\otimes  \mathbb{I}_2$ and  $B_0=E_1+E_2-E_3-E_4=\Gamma_{3} \, \sigma_3\otimes\sigma_3$, where $\Gamma_{1}=\frac{2ab(y_1-2xy_1-xy_2)}{y_1+y_2}$, $\Gamma_2=\frac{(1-2b^2)(2x-1)y_1+y_2(1-2b^2x)}{y_1+y_2}$ and $\Gamma_{3}=\frac{y_1-y_2+2xy_2}{y_1+y_2}$ with $a$, $b$,  $y_1$ and $y_2$ are defined earlier (Case-II) after Eq.(\ref{povm2}). It can be checked that the maximum eigenvalue of $B_0$ and $B_1$ is less than or equal to unity for all $\lambda \in [0,1]$ and for all $x \in [0,1]$. Hence, the conditions- $|\langle B_0\rangle|\leq1$ and $|\langle B_1\rangle|\leq1$ are satisfied and we can use the bilocality inequality (\ref{63}).


Now, to find out  the maximum magnitude of $\mathcal{B}_{bilocal}$ mentioned in Eq.(\ref{63}), we evaluate $I$ and $J$ following the similar process that was done in  case of the previous POVM. These can be written as 
\begin{eqnarray}
\label{82}
I&=&\mbox{Tr}\left[ \left\{ \left(\mathbf{u} + \mathbf{u}^{\prime} \right) \cdot \boldsymbol{\sigma} \otimes \Gamma_3 (\sigma_3 \otimes \sigma_3) \otimes \left(\mathbf{v}+\mathbf{v}^{\prime} \right) \cdot \boldsymbol{\sigma} \right\} \left(\rho_{AB}\otimes\rho_{BC} \right) \right]\nonumber\\
&=& \Gamma_3 \, \left(\sum_{i=1,2,3}(u_i+u_i^{\prime}) \,  t_{i3}^{AB} \right) \left( \sum_{k=1,2,3} (v_k+v_k^{\prime}) \, t^{BC}_{3k} \right) 
\end{eqnarray}
and
\begin{eqnarray}
\label{83}
J&=&\mbox{Tr}\Big[ \Big\{(\mathbf{u}-\mathbf{u}^{\prime}) \cdot \boldsymbol{\sigma} \otimes (\Gamma_2 \, \sigma_3 \otimes \mathbb{I}_2 + \Gamma_{1} \, \sigma_1 \otimes \sigma_1) \otimes (\mathbf{v}- \mathbf{v}^{\prime}) \cdot \boldsymbol{\sigma} \Big\} \nonumber \\
&& \hspace{6.1cm} (\rho_{AB}\otimes\rho_{BC}) \Big]\nonumber\\
&=& \Gamma_2 \, \left(\sum_{i=1,2,3}(u_i-u_i^{\prime})t_{i3}^{AB} \right) \left(\sum_{k=1,2,3}(v_k-v_k^{\prime})r_{C|k}^{BC} \right)\nonumber\\
&& \hspace{0.8cm} + \Gamma_1 \, \left(\sum_{i=1,2,3}(u_i-u_i^{\prime}) t_{i1}^{AB} \right) \left(\sum_{k=1,2,3}(v_k-v_k^{\prime}) t^{BC}_{1k} \right).
\end{eqnarray}

Now, we have to maximize the   bilocality inequality (\ref{63}) with respect to the unit vectors  $\mathbf{u}$, $\mathbf{u}^{\prime}$, $\mathbf{v}$,  and $\mathbf{v}^{\prime}$.  Here, we choose the above unit vectors  in three-dimensional space which is the most general case. Let $\mathbf{u}$= $(\sin{\alpha}\cos{\beta}$, $\sin{\alpha}\sin{\beta}$, $\cos{\alpha})$, $\mathbf{u}^{\prime}$ = $(\sin{\alpha^{\prime}}\cos{\beta^{\prime}}$, $\sin{\alpha^{\prime}}\sin{\beta^{\prime}}$, $\cos{\alpha^{\prime}})$, $\mathbf{v}$ = $(\sin{\gamma}\cos{\phi}$, $\sin{\gamma}\sin{\phi}$, $\cos{\gamma})$, and $\mathbf{v}^{\prime}$ = $(\sin{\gamma^{\prime}}\cos{\phi^{\prime}}$, $\sin{\gamma^{\prime}}\sin{\phi^{\prime}}$, $\cos{\gamma^{\prime}})$. It can be checked that the
the maximum value  of $\mathcal{B}_{bilocal}$ is obtained at $\alpha=\alpha^{\prime}$, $\beta=0$, $\beta^{\prime}=\pi$, $\gamma=\gamma^{\prime}$, $\phi=0$, $\phi^{\prime}=\pi$ and
$\cos{\alpha}=\cos{\gamma}=\sqrt{\frac{|\Gamma_{3}\tau_{1}^{AB}\tau^{BC}_1|}{|\Gamma_{1}\tau_{2}^{AB}\tau^{BC}_2|+|\Gamma_{3}\tau_{1}^{AB}\tau_1^{BC}|}}$.
And the maximum value is given by,
\begin{eqnarray}
\label{85}
    &&\mathcal{B}_{bilocal}^{max}=2\sqrt{|\Gamma_{1}\tau_{2}^{AB}\tau^{BC}_2|+|\Gamma_{3}\tau_{1}^{AB}\tau_1^{BC}|}
\end{eqnarray}
The above expression infers that when Alice-Bob and Bob-Charlie initially share two arbitrary two-qubit states between them produced from two independent sources and Bob performs the POVM given by (\ref{povm2}) on his qubits, then network nonlocality is demonstrated if
\begin{equation}
    |\Gamma_{1}\tau_{2}^{AB}\tau^{BC}_2|+|\Gamma_{3}\tau_{1}^{AB}\tau_1^{BC}|>1.
\end{equation}

In our scenario, we have considered maximally entangled Bell states: $\rho_{AB}=\rho_{BC}=|\phi^+\rangle\langle\phi^+|$. Hence, from (\ref{85}), the bilocality inequality becomes
\begin{equation}
    \label{86}
    \mathcal{B}_{bilocal}^{max}=2\sqrt{|\Gamma_{3}|+|\Gamma_{1}|}\leq2
\end{equation}
as $\tau_1^{AB}=\tau_2^{AB}=1$ and $\tau_1^{BC}=\tau_2^{BC}=1$ in the present case. Therefore, the bilocality inequality (\ref{63}) is violated when $|\Gamma_{33}|+|\Gamma_{11}|>1$.

Next, we find out the condition under which bilocality inequality (\ref{63}) is violated with the POVM given by Eq.(\ref{povm2}) with different values of $x$ using the expressions obtained above.  The results obtained are presented in  Table \ref{table6}. In particular, we have find out different  values of $x$ for which the network is non-bilocal or bilocal with respect to the quantum violation of the bilocality inequality (\ref{63}) for the whole range of the measurement parameter $\lambda \in [0,1]$. Interestingly, we observe that for $x=0.5$ the generated correlation does not violate the bilocality inequality (\ref{63}) for the whole range of the measurement parameter  $\lambda$.  However, the states shared between the pairs (1,4), (1,2) and (3,4) after completion of the generalized entanglement swapping protocol are entangled for the whole range of measurement parameter $\lambda$ except at $\lambda=0$ when the POVM (\ref{povm2}) with $x=0.5$ is performed by Bob. Also, for $x=0.725$, the generated correlation demonstrate network nonlocality for the whole range of $\lambda$ except at $\lambda=0$, whereas the states shared by the pairs (1,4), (1,2) and (3,4) after completion of the protocol do not violate the Bell-CHSH inequality (see Table \ref{tab3}) for any $\lambda$ in this case. In other words, after a generalized entanglement swapping protocol, one may end up with states that seem to be (Bell) local, while, in fact, they are (network) nonlocal.

\begin{table}
\begin{center}
\begin{tabular}{ |m{2cm}|m{2cm}| } 
 \hline
 x & Non-bilocal when  \\ 
 \hline
 x=0.3 & $0<\lambda<0.51$  \\
 \hline
  x=0.5 & Never \\
 \hline
 x=0.725 & $0<\lambda\leq1$ \\
 \hline
 x=0.8 & $0<\lambda\leq1$ \\
 \hline
\end{tabular}
\caption{The ranges of the measurement parameter $\lambda$ for which the network is non-bilocal or bilocal with respect to the quantum violation of the bilocality inequality (\ref{63})  when the POVM (\ref{povm2}) with different $x$ is performed by Bob.}
\label{table6}
\end{center}
\end{table}

\section{Concluding discussions} \label{sec5}

In the present study,  we have considered a generalized entanglement swapping protocol, where Alice and Bob share a Bell  pair ($1,2$) whereas Bob and Charlie share another Bell pair ($3,4$) in the same state. Bob performs a joint quantum measurement (POVM) on the pair ($2,3$) and discloses the outcome via classical communication to Alice and Charlie.  In this scenario, our key findings are twofold. At first, using generalised entanglement swapping protocol, we have created different kinds of quantum correlated single-parameter families of two-qubits mixed states in (1,4). These states have the following features: 

\begin{itemize}

\item a single parameter class of mixed states that gradually shows entanglement, EPR steering and Bell nonlocality as the state parameter is varied, 

\item a single parameter class of  mixed states that is entangled, but not EPR steerable (hence, not Bell nonlocal) for the whole range of the state parameter,

\item a single parameter class of  mixed entangled states that is steerable, but not Bell nonlocal for the whole range of the state parameter,

  \item single parameter class of mixed entangled states that is Bell nonlocal for the whole range of the state parameter.
\end{itemize}

 Secondly, in the standard entanglement swapping protocol,  the quantum correlation is completely transferred from the pairs (1,2) and (3,4) to the pair (1,4). Therefore, the natural question that arises in this context is how such transfer of quantum correlations changes when the standard entanglement swapping protocol is generalized by changing the initial states, or by changing the measurements, or both. \cite{sbose,gour,perseguers,perseguers2,nonlsw1,nonlsw2}. In the present study, we have addressed such questions by demonstrating non-trivial transfers of quantum correlations (e.g., Bell nonlocality, EPR steering, entanglement) from (1,2) and (3,4) to (1,4) by choosing different quantum measurements and/or initial states that are summarized below:

\begin{itemize}
    \item For the first POVM considered by us, we have shown that different quantum correlations are generated gradually (i.e., entanglement is generated at first, followed by the generation of EPR steering and then Bell nonlocality) in ($1,4$) with an increase in the measurement parameter at the expense of gradual destruction of these three quantum correlations in ($1,2$) and ($3,4$).
    
    \item For the second class of POVM, entanglement is generated in the pair ($1,4$) although the entanglement in each of the pairs ($1,2$) and ($3,4$) is not destroyed completely. Further, Bell-nonlocality or EPR steering in each of the pairs ($1,2$) and ($3,4$) is not  completely lost and, consequently, these two correlations are not generated in ($1,4$).
    
    \item In case of the third POVM, the pair ($1,4$) gains entanglement although the entanglement in each of the pairs ($1,2$) and ($3,4$) is not destroyed completely for whole range of the measurement strength $\lambda$ except at $\lambda=0$. Further, Bell-nonlocality is completely lost as a result of the protocol in each of the pairs ($1,2$) and ($3,4$) whereas Bell nonlocality is not generated in ($1,4$).
    
    \item For the fourth case of POVM, the pair ($1,4$) gains Bell nonlocality (and, hence, EPR steering as well as entanglement). On the other hand, each of the pairs ($1,2$) and ($3,4$) completely losses EPR steering and Bell nonlocality, but retains entanglement.

\end{itemize}

Hence, the present article provides with operational tools  in the basic entanglement swapping setup for generating different kinds of correlated states between two spatially separated observers without any interaction between them. In each of the above case, the whole class can be generated by varying the measurement parameter. Since the three types of quantum correlations, namely entanglement, EPR steering, Bell nonlocality, act as resources in  different kinds of tasks, each of the above mentioned classes of  mixed states has a distinct role as resource depending on the type of task to be performed. Remarkably, our results point out that all these types of quantum correlated  mixed states can be generated in a single setup just by choosing appropriate POVMs. We thus expect that our results will be helpful for experimental generation of different types of correlations in a single setup that can be employed in different information processing and communication tasks. 

 Note that experiments on entanglement swapping have been reported in several studies where  maximally entangled initial states were produced with high fidelity \cite{jin2014,zhang2019,liu2021,saunders2017}. However, in order to take into account of realistic non-idealness, we have further extended our studies by taking mixed initial states shared by the pairs (1,2) and (3,4). In such cases also, we have shown that the aforementioned different kinds of quantum correlated single-parameter families of two-qubits  states can be generated in (1,4) by choosing appropriate POVM for Bob.
 
 Here, we want to clarify that while studying standard Bell nonlocality or EPR steering of the states after completion of the generalized entanglement swapping protocol, we have not considered the possibility that the two sources producing the initial entangled states may be  independent. That is why we have also studied network nonlocality in the generalized entanglement swapping scenario considered by us as this concept incorporates the fact of independence of sources that is relevant in the context of entanglement swapping scenarios  \cite{brain2010,tavakoli2022}. Here, we have found out appropriate joint POVM performed by Bob on the pair (2,3)  for which the resulting statistics demonstrate or does not demonstrate network nonlocality for the whole range of the measurement parameter. Interestingly, we have shown an instance of POVM for which the measurement statistics does not show network nonlocality, but all the states shared by (1,2), (3,4) and (1,4) becomes entangled after the generalized entanglement swapping for the whole range of the measurement parameter.  We have also presented an example where all the states shared by the pairs (1,2), (3,4) and (1,4) after completion of the entanglement swapping protocol are Bell local if we consider that the sources producing the initial entangled states are not necessarily independent, while  they show network nonlocality when the independence of sources has been taken into account. This may hold for EPR steering as well.

Before concluding we would like to mention some future research directions that we leave open. Considering multipartite entanglement swapping protocol \cite{sbose} assisted with different kinds of POVM for generating different kinds of genuine multipartite quantum correlations  is worth to be studied in future. 
 Also, probing network steering (incorporating statistical independence of the sources) in the context of our work is another direction for future research. Finally, finding out the amount of information loss or gain by different pairs \cite{lg1,pbej} in our entanglement swapping setups is of fundamental interest.

\section*{Acknowledgements}
The authors are grateful to Prof.
Somshubhro Bandyopadhyay for helpful discussions. SM acknowledges the Ministry of Science and Technology, Taiwan (Grant No. MOST 110-2124-M-002-012). DD acknowledges the Science and Engineering Research Board (SERB), Government of India for financial support through a National Post Doctoral Fellowship (File No.: PDF/2020/001358). During the later phase of this work, the research of DD is supported by the Royal Society (United Kingdom) through the Newton International Fellowship (NIF$\backslash$R1$\backslash$212007).

\end{document}